\documentclass[Times,twocolumn,tighten]{aastex63}

\submitjournal{ApJ}

\shorttitle{Accretion Dynamics of MAXI J1348-630}
\shortauthors{A. Jana et al.}

\begin{document}

\title{Accretion Flow Evolution of a New Black Hole Candidate MAXI J1348-630 During the 2019 Outburst}

\correspondingauthor{Arghajit Jana}
\email{argha0004@gmail.com}

\author[0000-0001-7500-5752]{Arghajit Jana}
\affiliation{Physical Research Laboratory, Navrangpura, Ahmedabad 380009, India}
\affiliation{Indian Center for Space Physics, 43 Chalantika, Garia St. Road, Kolkata 700084, India}

\author[0000-0003-1856-5504]{Dipak Debnath}
\affiliation{Indian Center for Space Physics, 43 Chalantika, Garia St. Road, Kolkata 700084, India}

\author[0000-0001-6770-8351]{Debjit Chatterjee}
\affiliation{Indian Center for Space Physics, 43 Chalantika, Garia St. Road, Kolkata 700084, India}

\author[0000-0002-6252-3750]{Kaushik Chatterjee}
\affiliation{Indian Center for Space Physics, 43 Chalantika, Garia St. Road, Kolkata 700084, India}

\author[0000-0002-0193-1136]{Sandip Kumar Chakrabarti}
\affiliation{Indian Center for Space Physics, 43 Chalantika, Garia St. Road, Kolkata 700084, India}

\author[0000-0003-2865-4666]{Sachindra Naik}
\affiliation{Physical Research Laboratory, Navrangpura, Ahmedabad 380009, India}

\author[0000-0002-7658-0350]{Riya Bhowmick}
\affiliation{Indian Center for Space Physics, 43 Chalantika, Garia St. Road, Kolkata 700084, India}

\author[0000-0003-0071-8947]{Neeraj Kumari}
\affiliation{Physical Research Laboratory, Navrangpura, Ahmedabad 380009, India}


\begin{abstract}

Galactic black hole candidate MAXI~J1348-630 was recently discovered by {\it MAXI} and {\it Swift}/BAT satellites 
during its first outburst in 2019 January which continued for $\sim4$ months. We study the spectral and timing 
properties of the source in detail. The combined $1-150$ keV {\it Swift}/XRT, {\it Swift}/BAT and {\it MAXI}/GSC 
spectra are investigated with the two component advective flow (TCAF) solution. Physical flow parameters of TCAF, 
such as the Keplerian disk accretion rate, the sub-Keplerian halo accretion rate, the shock location and the shock 
compression ratio are estimated from our spectral fits. Based on the variation of flux in soft and hard X-ray 
ranges, the hardness ratio, TCAF model fitted accretion rates and accretion rate ratio (ARR), we show how the 
sources evolved through four spectral states viz. hard, hard-intermediate, soft-intermediate and soft states 
in rising and declining states. Low-frequency quasi-periodic oscillations (QPO) are observed in two observations 
during the rising phase of the outburst. From the spectral analysis, we estimate the mass of the black hole 
to be $9.1^{+1.6}_{-1.2}$ $M_{\odot}$. We also find that the viscous timescale in this outburst is $\sim 3.5$ days. 
The distance of the source is also estimated as $5-10$ kpc from state transition luminosity.

\end{abstract}


\section{Introduction}
Black hole X-ray binaries (BHXRBs) consist of a black hole (BH) and a companion main-sequence star. 
Transient BHXRBs spend most of the time in the quiescent state. They occasionally go into outbursts, 
which often last from weeks to months. During an outburst, the X-ray intensity of the source could
rise by several orders of magnitude in comparison to that in the quiescence phase. In a BHXRB, 
matter from the companion star is accreted onto the central black hole and forms an accretion
disk. In the accretion process, the gravitational potential energy of the accreted matter is converted
to heat which is radiated in the entire electromagnetic wavebands, e.g., from radio to $\gamma$-ray 
range. An outbursting BHXRBs show rapid changes in both spectral and temporal properties.
The hardness-intensity diagram (HID; Belloni et al. 2005) or accretion rate-intensity diagram (ARRID; 
Jana et al. 2016) show the correlation between spectral and temporal parameters in different spectral
states. In general, a BHXRB exhibits four different spectral states, namely,  hard state (HS),
hard-intermediate state (HIMS), soft-intermediate state (SIMS), and soft state (SS). When a BHXRB
evolves through these states in the sequence HS $\rightarrow$ HIMS $\rightarrow$ SIMS $\rightarrow$
SS $\rightarrow$ SIMS $\rightarrow$ HIMS $\rightarrow$ HS, it produces a so-called hysteresis loop 
(see for more details, Remillard \& McClintock 2006, Debnath et al. 2013, and reference therein).
The BHXRBs also exhibit quasi-periodic oscillations (QPOs) in some spectral states (see Remillard 
\& McClintock 2006 for a review). Unlike the high-frequency QPOs (HFQPO; QPO frequency $\geq$40 Hz) 
which are rare, low-frequency QPOs (LFQPOs) are common in BHXRBs and are classified into three types
- A, B, and C (Casella et al. 2005), depending on their nature (Q-value, RMS amplitude, noise, etc.). 
Different spectral and temporal properties characterize each spectral state. In the HS and HIMS, 
the hard X-ray flux dominates with evolving type-C QPOs. In the SIMS and SS, the soft X-ray flux 
dominates over the hard X-ray flux. Type-A or type-B QPOs may be observed sporadically in the SIMS. 
QPOs are not seen in the SS.

In general, an X-ray spectrum of BHXRBs consists of two components: a soft multicolor disk blackbody 
(diskbb) and a hard power-law (PL) component. The multicolor black body component is originated from 
a standard thin disk (Novikov \& Thorne 1973; Shakura \& Sunyaev 1973), while the power law is originated 
from a hot Compton cloud consisting of hot electrons (Sunyaev \& Titarchuk 1980, 1985). There exist 
various models in the literature to explain the nature of the Compton cloud, e.g., magnetic corona 
(Galeev et al. 1979), evaporated disk (Esin et al. 1997), disk-corona model (Zdziarski et al. 1993), 
two component advective flow (TCAF) solution (Chakrabarti \& Titarchuk 1995, hereafter CT95; 
Chakrabarti 1997), etc. Except for TCAF, which is based on viscous and radiative transonic flow solutions, 
other models are phenomenological.

In TCAF, the accretion disk has two components: an optically thick, geometrically thin high viscous 
Keplerian flow on the equatorial plane submerged inside an optically thin, low viscous sub-Keplerian 
flow (for a review on TCAF, see Chakrabarti 2018). The sub-Keplerian flow moves faster and temporarily
slow down at the centrifugal barrier and forms an axisymmetric shock (Chakrabarti, 1990) when the 
inflowing matter piles up at the barrier. The post-shock region being hot and puffed up acts as the 
Compton cloud and is known as CENtrifugal pressure supported BOundary Layer (CENBOL). The soft photons
which are originated from the Keplerian disk, contribute to the multicolor blackbody spectrum. A 
fraction of these photons is intercepted by the CENBOL and undergoes inverse-Compton scattering with
the hot electrons of the CENBOL and become hard photons. These hard photons produce the hard power law 
tail observed in the spectra. Physical oscillation of the CENBOL causes the fraction of intercepted photons
to oscillate, resulting in LFQPOs often observed in the power density spectra (PDS). CENBOL oscillation
is triggered when the compressional heating time scale of the flow roughly matches the radiative cooling
time scale and a resonance condition is satisfied (Molteni et al. 1996). It may also be triggered if 
Rankine-Hugoniot conditions are not satisfied in a time-dependent transonic flow even though there are 
two physical sonic points (Ryu et al. 1997). A bipolar jet is launched from the hot CENBOL region in 
the harder states (Chakrabarti, 1999). The jet is absent when the CENBOL itself is collapsed in the softer states. 

In general, an outburst is believed to be triggered by the sudden rise of viscosity at the outer 
edge of the disk (Ebisawa et al. 1996). When an outburst starts, the sub-Keplerian flow rushes
towards the BH and forms an axisymmetric shock at the centrifugal boundary. A strong shock is
formed with a large and hot CENBOL. In the initial phase, the Keplerian disk accretion rate is
low (as it moves in viscous time scale), and therefore, it can not cool the CENBOL efficiently, 
and hard state is observed. In this state, ARR is found to decrease with the progress of the 
outburst as we see a monotonic rise in the disk accretion rate. The shock becomes weak as it 
moves towards the BH. A strong, compact jet could be observed in this state of the outburst. 
The HS is associated with the evolving type-C QPOs. The QPO frequency monotonically increases 
with the progress of the outburst as the shock location decreases. This is because, the QPO 
frequency ($\nu$) varies with the shock location as $\nu \sim X_s^{-3/2}$ (Chakrabarti \& 
Manickam 2000; Chakrabarti et al. 2008). QPOs exist as long as the resonance condition due 
to rough agreement between the heating and cooling of the CENBOL is satisfied. 

As the outburst progresses, the source enters into the HIMS. In this state, the Keplerian disk 
accretion rate becomes comparable with the sub-Keplerian halo accretion rate. As a result, ARR 
further decreases. The shock continues to move inward. Evolving type-C QPO is observed in this 
state as well. With the further rise in the Keplerian disk accretion rate, the BH exhibits SIMS.
Here, the disk accretion rate is comparatively higher than the sub-Keplerian halo accretion rate.
A discrete ejection or blobby jet could be observed in this state. Sporadic type-B or A QPOs could
be found in this state. The shock becomes weak and moves inward. With the progress in the outburst,
the source enters into the SS. In this state, the Keplerian disk accretion rate efficiently cools 
down the CENBOL. No QPO is observed in this state. Generally, we do not see any jet in this spectral 
state.  

The source enters the declining phase when the viscosity is turned off or reduced (Ebisawa et al.
1996, Roy \& Chakrabarti, 2017). As the Keplerian disk is already formed, it is difficult to drain 
the matter as the viscosity is reduced. Thus, the source remains in the SS and the SIMS in the 
declining phase for a relatively long time. As the outburst progress further, the accretion rate 
decreases and the source goes through the HIMS and the HS before going to the quiescence state.
In the declining phase, evolving decreasing QPO frequency is observed in the HIMS and HS. One
could also observe outflows in these two harder spectral states.

To fit a spectrum, TCAF uses only four flow parameters, namely, accretion rates of the Keplerian
and the sub-Keplerian components, the size of the CENBOL (i.e., the shock location) and density 
variation inside CENBOL required to obtain the optical depth (obtained from the shock strength). 
Apart from these, one instrument parameter, namely the Normalization factor (which gives the 
ratio of emitted to observed photon spectrum) and one system parameter, namely, the mass of
the black hole are required. In 2014, the TCAF solution was implemented in {\tt XSPEC} (Arnaud 1996) 
as a local additive model to carry out spectral analysis of black hole X-ray binaries (Debnath et 
al. 2014, 2015a for more details). The accretion dynamics of several black hole candidates (BHCs)
are studied successfully with this model (Mondal et al. 2014, 2016; Debnath et al. 2015b, 2017, 
2020; Jana et al. 2020b; Chatterjee et al. 2019, 2020; Molla et al. 2017; Shang et al. 2019).
In each case, we have obtained the evolution of the actual physical parameters of the flow as 
well as the mass of the black hole very successfully. This motivated us to study the properties
of the newly discovered BHC MAXI~J1348-630 during the 2019 outburst with the TCAF model.

MAXI~J1348-630 is a Galactic BHXRB which was discovered very recently on 2019 January 26 by 
{\it MAXI}/GSC (Yatabe et al. 2019) at R.A. = $13^h$ $48'$ $12''$, DEC = $-63^{\circ}$ $4'$ $4''$.
Later, the {\it Swift}/XRT localized the position of the source at R.A. = $13^h$ $48'$ $12''.73$, 
DEC = $-63^{\circ}$ $16'$ $26''.8$ (Kennea et al. 2019). The source was also observed by {\it INTEGRAL},
{\it NICER} and {\it HXMT} (Lepingwell et al. 2019, Sanna et al. 2019, Chen et al. 2019). Optical
(Denisenko et al. 2019, Russell et al. 2019a) and radio (Russell et al. 2019b) observations of the 
source were also carried out during the outburst. The outburst lasted for about four months. 
MAXI~J1348-630 was observed to re-brighten a few times after this main outburst (Negoro et al. 2019). 
Observation of QPO on 2019 January 30 was reported with the {\it Swift}/XRT and {\it HXMT} observation
(Jana et al. 2019; Chen et al. 2019). A preliminary data analysis with the TCAF model leads us to 
estimate the mass of the black hole to be in the range of $8.5 - 11$~$M_{\odot}$ (Jana et al. 2019). 
However, Tominaga et al. (2020) estimated the mass of the black hole as 16 $M_{\odot}$ from the 
spectral analysis with the {\it MAXI} data. They have also estimated the distance of the source 
as $4-8$~kpc.

In this paper, we study the accretion flow dynamics of MAXI~J1348-630 with the TCAF model. 
The paper is organized in the following way. In \S 2, we discuss observation and data analysis. 
The results obtained from this work are presented in \S 3. In \S 4, we present discussion and 
concluding remarks.


\section{Observation and Data Analysis}

We studied the newly discovered MAXI~J1348-630 during its 2019 outburst using {\it Swift} and 
{\it MAXI} data in $1-150$ keV energy range. In our analysis, we used a total of 27 observations
of the source with the {\it Swift}/XRT ($1-10$~keV range), {\it Swift}/BAT ($15-150$ keV range)
and {\it MAXI}/GSC ($7-20$ keV range) between 2019 January 26 and 2019 May 15. Among 27 observations,
there were four epochs of observations during which all three instruments, eleven epochs of observations
during which {\it Swift}/XRT and {\it MAXI}/GSC and one epoch of observation during which {\it Swift}/XRT 
and {\it Swift}/BAT were used simultaneously. There were one and ten epochs of observations during which 
only {\it Swift}/BAT and {\it Swift}/XRT were used, respectively. The details of the log of observations 
of the source used in the present work are given in Table 1. 

We used WT mode data for the XRT observations.  We used grade-0 data to reduce pile-up effects. Using
the {\tt xrtpipeline} command, cleaned event files were generated. A circular region of a radius of 30 
pixels was chosen for the source. Using {\tt XSELECTv2.4} task, source and background light curves (0.01 s
time resolution) and spectra were generated. We re-binned the spectra with 10 counts/bin using {\tt grppha} 
task. To generate {\it Swift}/BAT spectra, standard procedures were followed, as suggested by the 
instrument team. Using {\tt batbinevt} task, detector plane images (dpi) were generated. We used 
{\tt batdetmask} for appropriate detector quality. We ran {\tt bathotpix} to find noisy detector pixels 
and quality mapping. The {\tt batmaskwtevt} task was used to apply mask weighting to the event mode data. 
With {\tt batphasyserr}, a systematic error was applied to the BAT spectra. Using {\tt batupdatephakw} 
task, ray-tracing was corrected. The response matrices for the BAT spectra were generated with {\tt batdetmask}
task. The $7-20$ keV {\it MAXI}/GSC spectra were generated by the MAXI on-demand process web tool 
(Matsuoka et al. 2009). 

For spectral analysis, we use TCAF model-based fits file as the additive table model. As mentioned in the 
Introduction, to fit using TCAF, four input flow parameters, such as, the Keplerian disk accretion rate
($\dot{m}_d$ in $\dot{M}_{Edd}$), the sub-Keplerian halo accretion rate ($\dot{m}_h$ in $\dot{M}_{Edd}$), 
the shock location ($X_s$ in Schwarzschild radius $r_s$=$2GM_{BH}/c^2$) and the dimensionless shock compression 
ratio ($R = \rho_+ / \rho_-$, ratio of post-shock matter density to the pre-shock matter density) are essential.
Also, one system parameter, i.e., the mass of the black hole ($M_{BH}$ in $M_{\odot}$) and one instrument parameter, 
namely, the Normalization constant ($N$) are required. In case $M_{BH}$ is not known, we can extract it as well.
Each observation yields a best-fitted value of mass along with other parameters (see Appendix).
The average of these masses comes out to be $ 9.1$ $M_{\odot}$. Freezing the mass at this value, we re-analyze
all the data to obtain the final value of each parameter.

We ran {\tt powspec} task to generate the power density spectra (PDS) from the $0.01$~sec time binned 
XRT light curves in the energy range of $1-10$~keV. We searched for the presence of low-frequency QPOs 
in the PDS. We studied the PDS with a different number of bins (i.e., 1024, 2048, 4096 \& 8192) and 
subintervals. To obtain frequency ($\nu$), width ($\Delta \nu$), Q-value ($\nu/\Delta \nu$), RMS amplitude,
etc., we fitted observed QPOs with the Lorentzian model. To fit the continuum of the PDS, we used power law
and linear models with the Lorentzian model.

\section{Results}

\begin{figure*}
\vskip 0.2cm
\centering
\includegraphics[width=12cm,keepaspectratio=true]{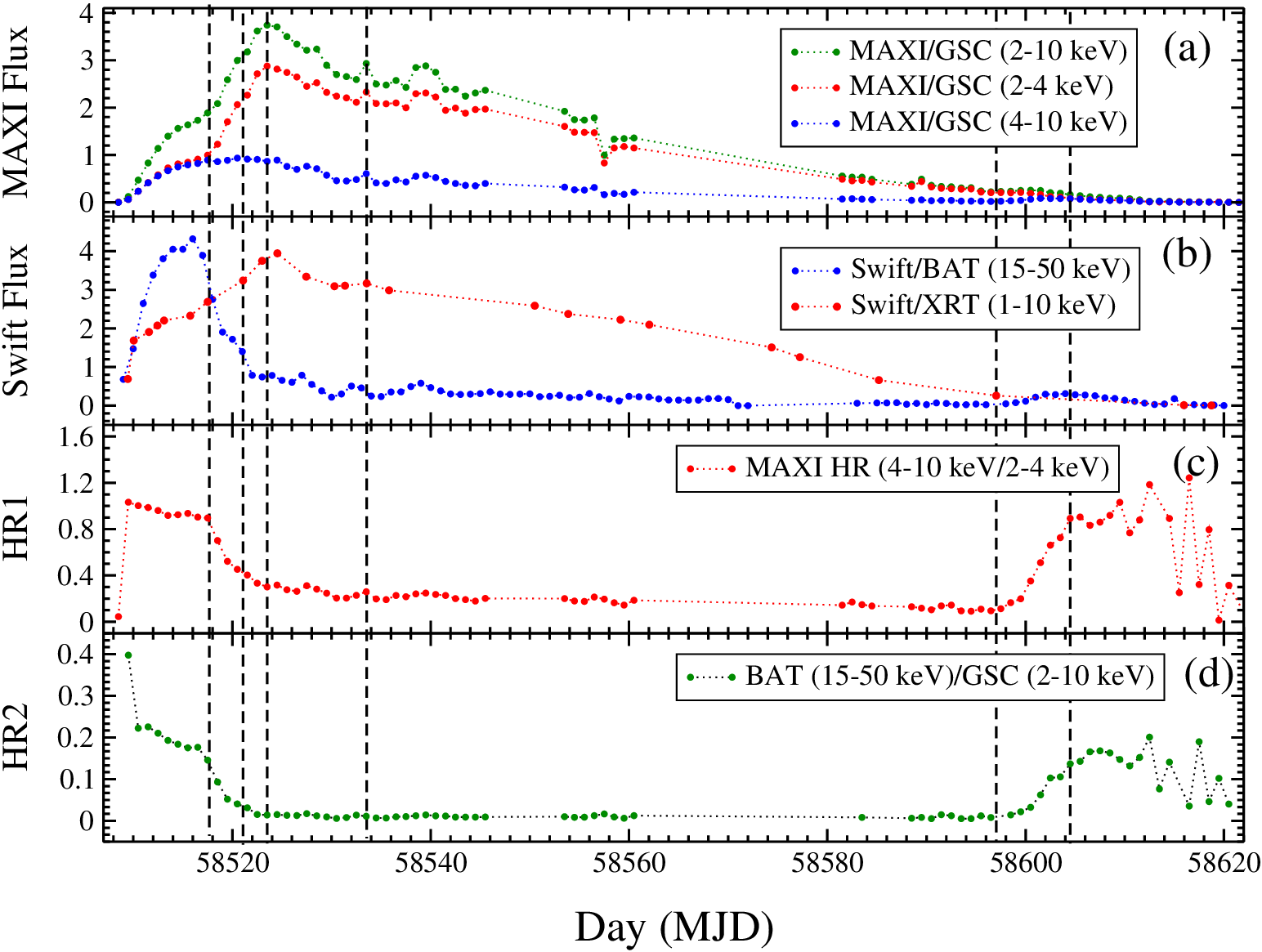}
\caption{The variation of (a) {\it MAXI}/GSC flux in $2-10$ keV, $2-4$ keV and $4-10$ keV in the unit of $Crab$, 
(b) $1-10$~keV {\it Swif}/XRT flux and $15-150$ keV {\it Swift}/BAT flux in the unit of $Crab$, (c) hardness ratio 
e.g. ratio between  $4-10$ keV and $2-4$ keV flux from {\it MAXI}/GSC data (HR1) and (d) ratio between flux in 
$15-50$ keV (from {\it Swift}/BAT data) and $2-10$ keV (from {\it MAXI}/GSC data) ranges (HR2) are shown with time 
(in MJD). The verticle lines mark spectral state transition days. The source evolved in the sequence of HS
$\rightarrow$ HIMS $\rightarrow$  SIMS $\rightarrow$  SS $\rightarrow$ SIMS $\rightarrow$ HIMS $\rightarrow$ HS.
\label{profile}}
\end{figure*}

\subsection{Outburst Profile}

MAXI J1348-630 was in the outbursting phase for almost four months. The light curves in different energy
ranges and the hardness ratios during the entire outburst, are shown in several panels of Figure~\ref{profile}. 
From the {\it MAXI} and {\it Swift}/XRT light curves (panels (a) and (b) of Figure~\ref{profile}), the 
outburst can be characterized as `$slow-rise-slow-decay$' (SRSD, Debnath et al. 2010). In Figure~\ref{profile}
(a), we show the variation of {\it MAXI}/GSC flux in three different energy bands e.g., $2-10$~keV, $2-4$ 
keV and $4-10$ keV ranges during the outburst. The variation of $1-10$ keV {\it Swift}/XRT flux and $15-50$ 
keV {\it Swift}/BAT flux covering the entire duration of X-ray outburst, are shown in the second panel 
(panel-b). The hardness ratio-1 (HR1)-- ratio between fluxes in $4-10$ keV and $2-4$ keV ranges (obtained 
from {\it MAXI}/GSC) and hardness ratio-2 (HR2) -- ratio between $15-50$ keV flux (obtained from {\it Swift}/BAT)
and $2-10$ keV flux (obtained from {\it MAXI}/GSC) are shown in the third panel (panel-c) and the bottom panel
(panel-d) of Figure~\ref{profile}, respectively.

From Figure~\ref{profile}(a), we see that the total flux ($2-10$ keV {\it MAXI}/GSC flux) and the soft 
X-ray flux ($2-4$ keV {\it MAXI}/GSC flux) increased slowly from 2019 January 25 (MJD 58508) and reached
the maximum value on 2019 February 9 (MJD 58523). After that, the flux in both the energy ranges decreased
gradually. The $4-10$ keV {\it MAXI}/GSC flux also increased slowly and attained maximum on 2019 February 6 
(MJD 58520), which was three days before the peak of the soft X-ray ($2-4$ keV) flux. After that, it 
decreased slowly till the end of the outburst. High energy ($15-50$~keV range) {\it Swift}/BAT flux increased
rapidly compared to the {\it MAXI}/GSC flux. It attained its peak on MJD 58516, after which it declined sharply
until 2019 February 8 (MJD 58522). After the sharp decline, the BAT flux decreased very slowly. The BAT flux
again increased briefly since 2019 April 27 (MJD 58600) though soon after, it declined again from MJD 58605. 
The $1-10$ keV {\it Swift}/XRT fluxes were estimated from the spectral fitting by using the TCAF model. 
From the second panel of Figure~\ref{profile}, it can be seen that the XRT flux increased slowly at the 
rising phase and was maximum on 2019 February 10 (MJD 58524.53). It is possible that if the XRT observation
were available, the XRT flux would have attended a maximum value on the 2019 February 9 (MJD 58523) when
the $2-10$ keV flux from {\it MAXI}/GSC was maximum. The {\it Swift}/XRT flux decreased slowly after 
reaching the maximum value.

\subsection{Hardness Ratio}

At the start of the outburst, both hardness ratios (HR1 and HR2) were high. As the outburst progressed, 
they decreased slowly till 2019 February 3 (MJD 58517), after which both HR1 and HR2 fell sharply. Beyond
2019 February 6 (MJD 58520), both HR1 and HR2 remained almost constant till 2019 April 27 (MJD 58600).
After that, both the hardness ratios increased as the source entered in the quiescence state. 

From the evolution of hardness ratios and X-ray fluxes in different energy bands of GSC and BAT, 
we have a rough idea about the spectral nature of the source. We found four spectral states, namely, 
HS, HIMS, SIMS, and SS as in classical or type-I transient BHCs (see, Nandi et al. 2012; Debnath et al.
2013). When the outburst started, the HRs were high and roughly constant until 2019 February 3 (MJD 58517.5). 
The source remained in the HS (ris.) until then from the start of the outburst. After that day, HRs 
decreased rapidly due to the rapid rise of the soft X-ray flux. So, the source entered in the HIMS (ris.). 
On 2019 February 7 (MJD 58521.5), the HRs started to fall slowly, and the source entered in the SIMS (ris.).
The $2-4$ keV {\it MAXI}/GSC and $1-10$ keV {\it Swift}/XRT flux showed a rapid increase in this state
compared to a roughly constant flux of the $4-10$ keV {\it MAXI}/GSC and decreasing $15-50$ keV {\it Swift}/BAT
flux. From 2019 February 9 (MJD 58523.5), both the $2-4$~keV and $2-10$~keV {\it MAXI}/GSC fluxes were 
at their maxima, and the source entered in the SS. Hardness ratio (HR) became almost constant (at low values)
throughout the SS. The transition from the SS to the SIMS cannot be marked with the variation of the HRs. The
HRs started to increase on 2019 April 24 (MJD 58597.5), and the source entered in the HIMS (dec.). 
On 2019 May 1 (MJD 58604.5), the HRs became roughly constant, and the source entered in the HS (dec.).

\begin{figure}
\vskip 0.2cm
\centering
\includegraphics[width=7.0cm,angle=270,keepaspectratio=true]{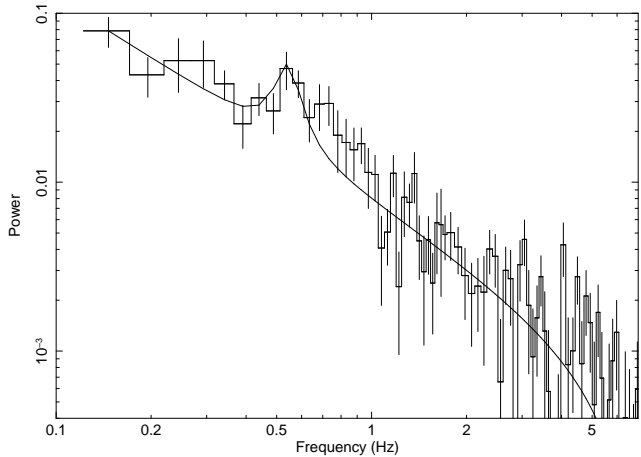}
\caption{Power density spectrum (PDS) of the source obtained from the {\it Swift}/XRT observation 
on 2019 January 29 (Observation ID = 00886496000). Quasi-period osscilations (QPO) at a 
frequency of $0.57$ Hz can be seen in the PDS.
\label{pds}}
\end{figure}

\subsection{Power Density Spectra}

We studied the power-density spectra (PDS) generated from the lightcurves in $1-10$ keV range with a time 
resolution of $0.01$ s, obtained from the {\it Swift}/XRT data. During our investigation of the presence
of QPO in the XRT lightcurves, we observed QPO only in 2 observations in the rising phase of the outburst 
e.g. on 2019 January 29 and 2019 January 30. The PDS are fitted with combined Lorentzian, linear, and power 
law models. The model fitted $\chi^2/dof$ obtained are 143/113 and 136/112 for two QPO observations on 2019
January 29 and 2019 January 30, respectively. In Figure~\ref{pds}, we show a power-density spectrum (PDS) 
observed on 2019 January 29 (MJD 58512.43), where a $0.57\pm0.04$ Hz QPO with Q-value of $4.5\pm0.4$ and 
$8 \pm 0.8 $\% rms was seen. The second QPO was detected at $0.66\pm 0.04$ Hz with Q-value of $2.8\pm0.3$ 
and $5 \pm 0.6$\% rms on 2019 January 30 (MJD 58513.11). We did not detect any QPO in the declining phase 
of the outburst. The rapid change in Q-values indicates a rapid falling out of the resonance condition, 
which is believed to be the origin of these QPOs.

\begin{figure*}
\vskip 0.2cm
\centering
\vbox{
\includegraphics[width=7.5truecm]{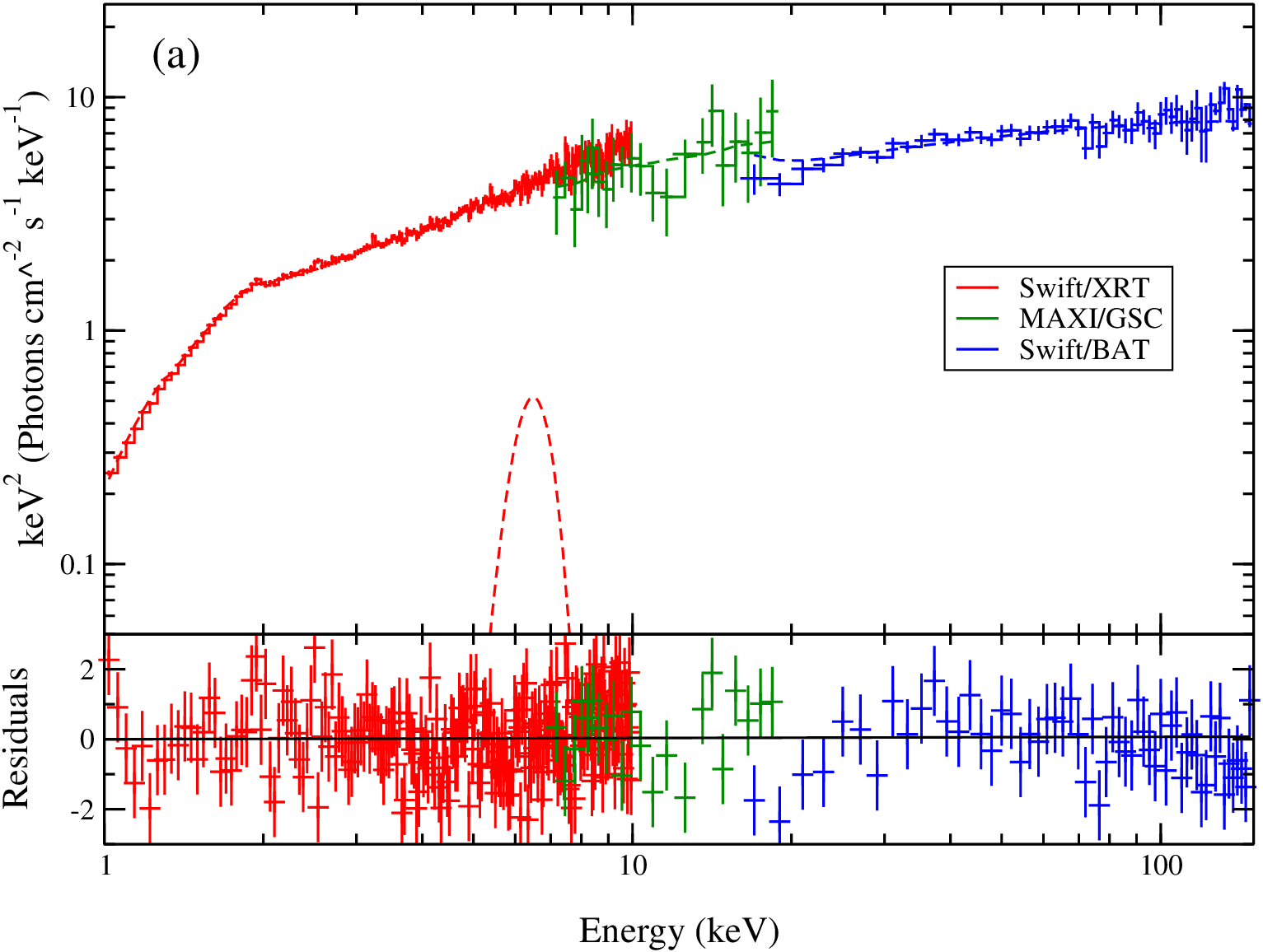}\hskip 0.5cm
\includegraphics[width=7.5truecm]{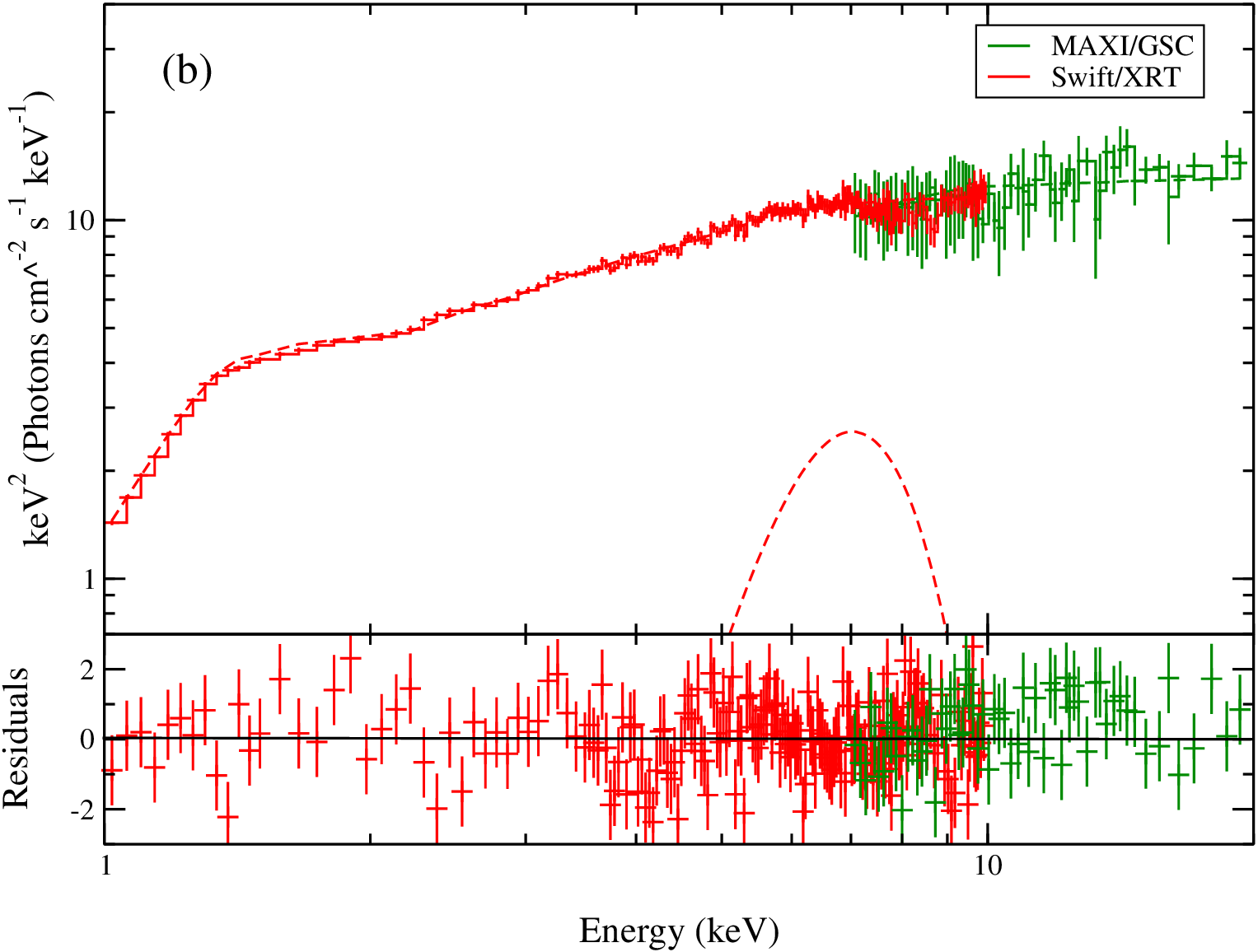}}
\hskip 0.2cm
\vbox{
\includegraphics[width=7.5truecm]{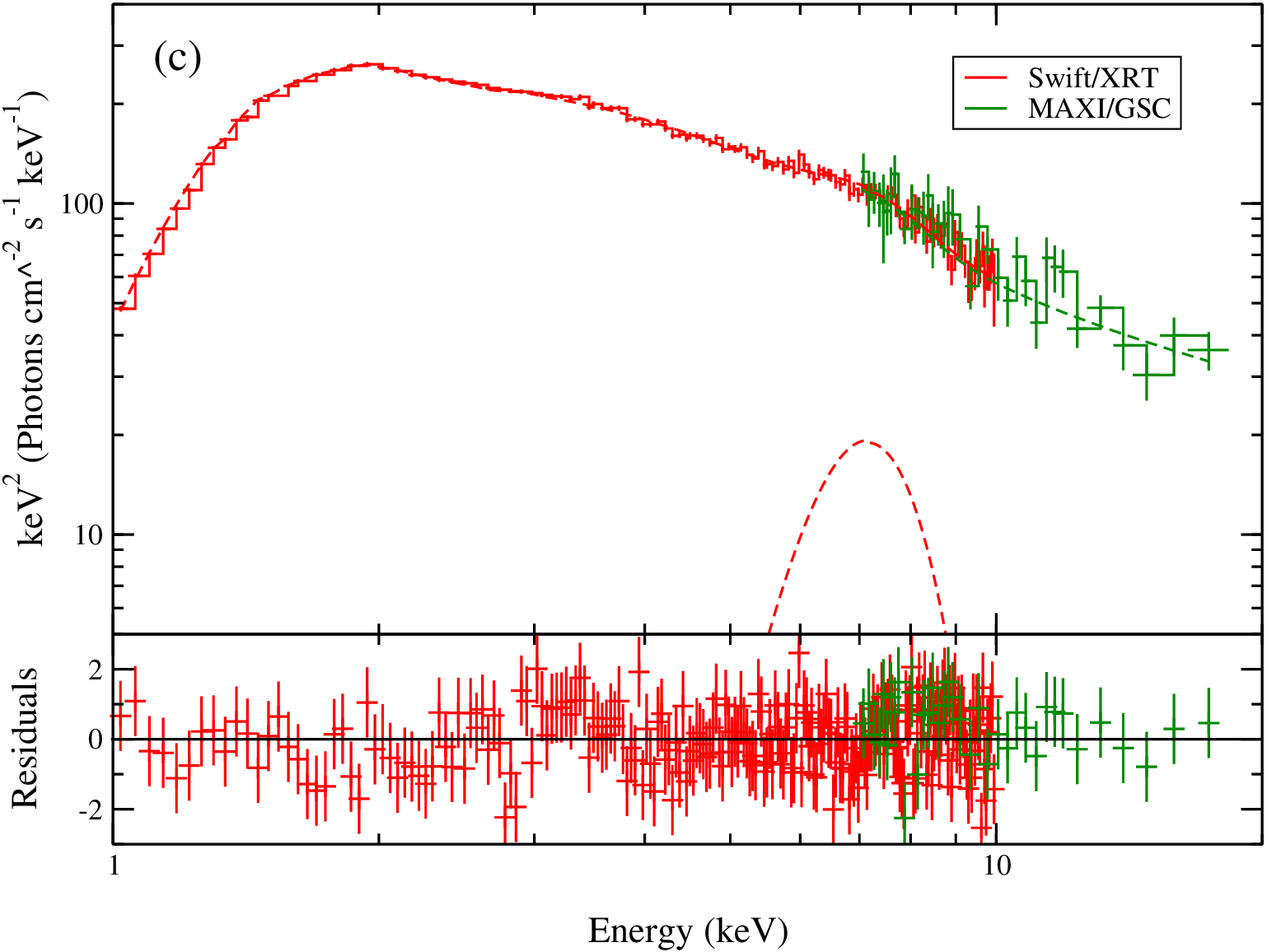}\hskip 0.5cm
\includegraphics[width=7.5truecm]{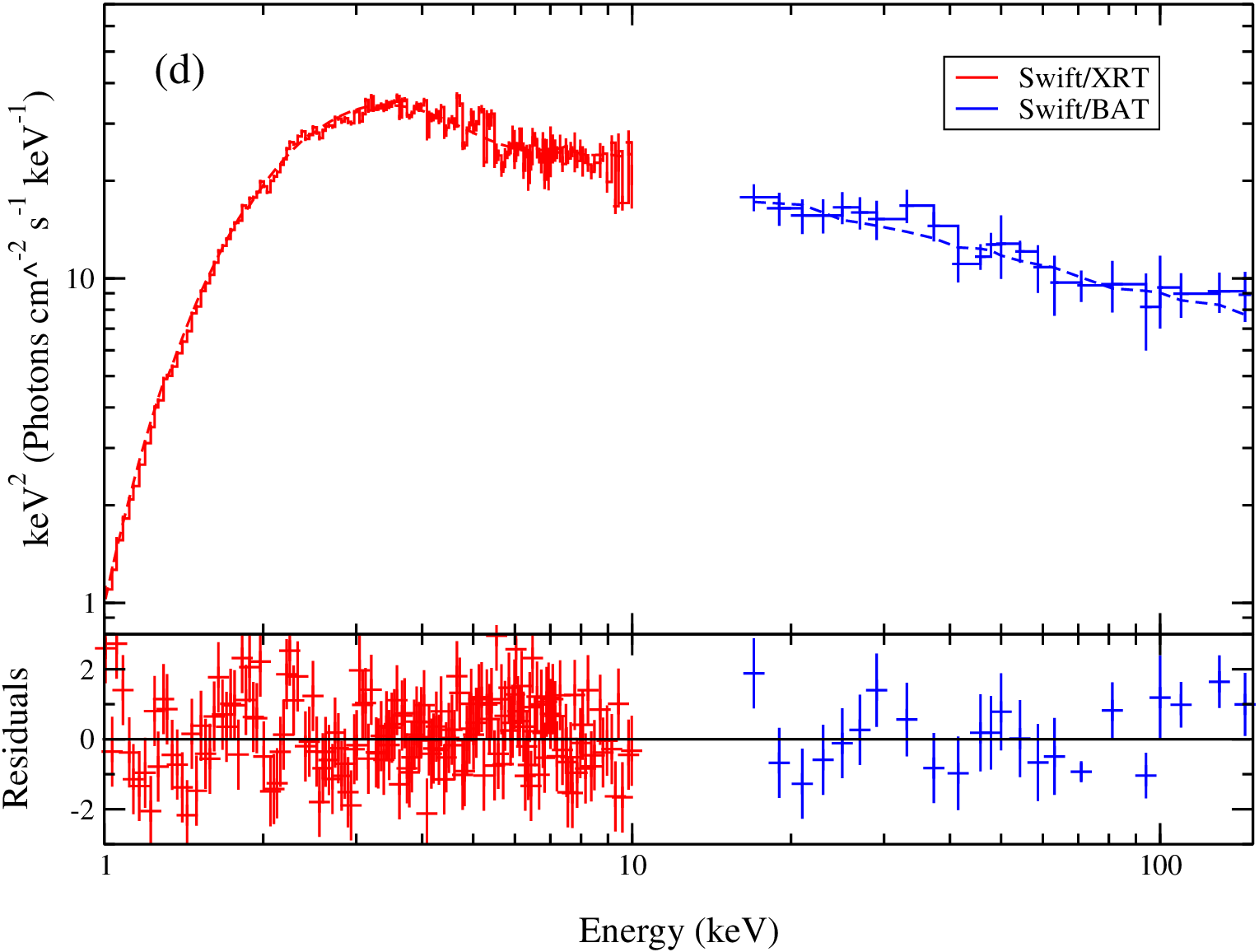} }
\vskip 0.1cm
\caption{TCAF model fitted spectra in four different states : (a) HS in the rising phase (Obs. ID - 00885960000, 
2019 January 27), (b)  HIMS in the rising phase (Obs. ID - 00011107002, 2019 February 03), (c) SS (Obs. ID - 
00011107007, 2019 February 10), and (d) SIMS in the declining phase (Obs. ID - 00896552000, 2019 April 04), 
respectively, are shown in the top panels along with the contribution to the residuals in the bottom panels.
\label{spectra}}
\end{figure*}

\subsection{Spectral Properties}

\begin{figure*}
\vskip 0.2cm
\centering
\includegraphics[width=13cm,keepaspectratio=true]{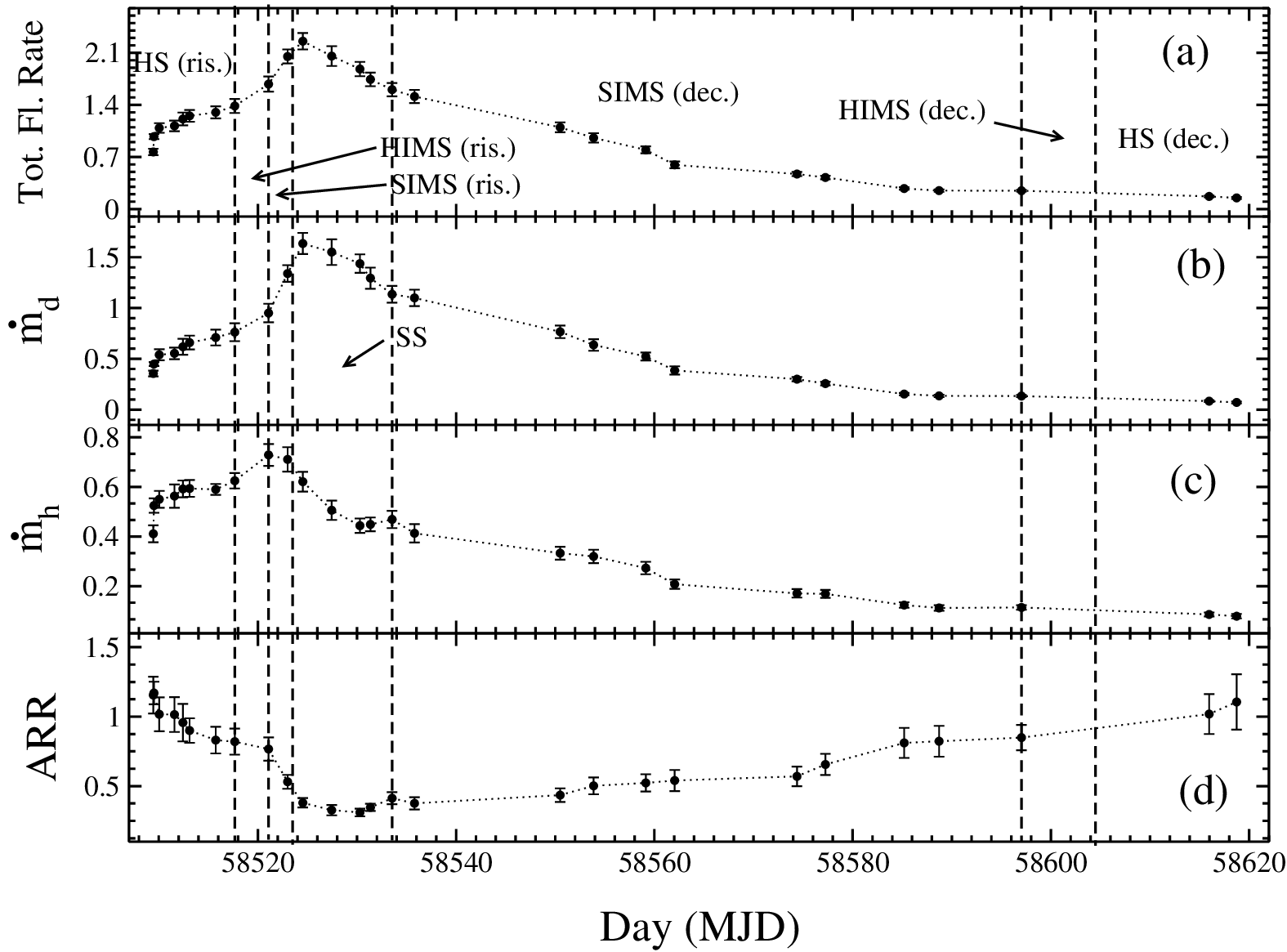}
\caption{The variation of (a) total accretion rate ($\dot{m}_d$ + $\dot{m}_h$ in $\dot{M}_{Edd}$),
(b) the Keplerian disk rate ($\dot{m}_d$ in $\dot{M}_{Edd}$), (c) the sub-Keplerian halo rate ($\dot{m}_h$ 
in $\dot{M}_{Edd}$) and (d) the accretion rate ratio (ARR = $\dot{m}_{h}$/$\dot{m}_{d}$) with time (MJD) 
during the outburst are shown. The verticle lines mark spectral state transition days.
\label{4}}
\end{figure*}

\begin{figure*}
\vskip 0.2cm
\centering
\includegraphics[width=13cm,keepaspectratio=true]{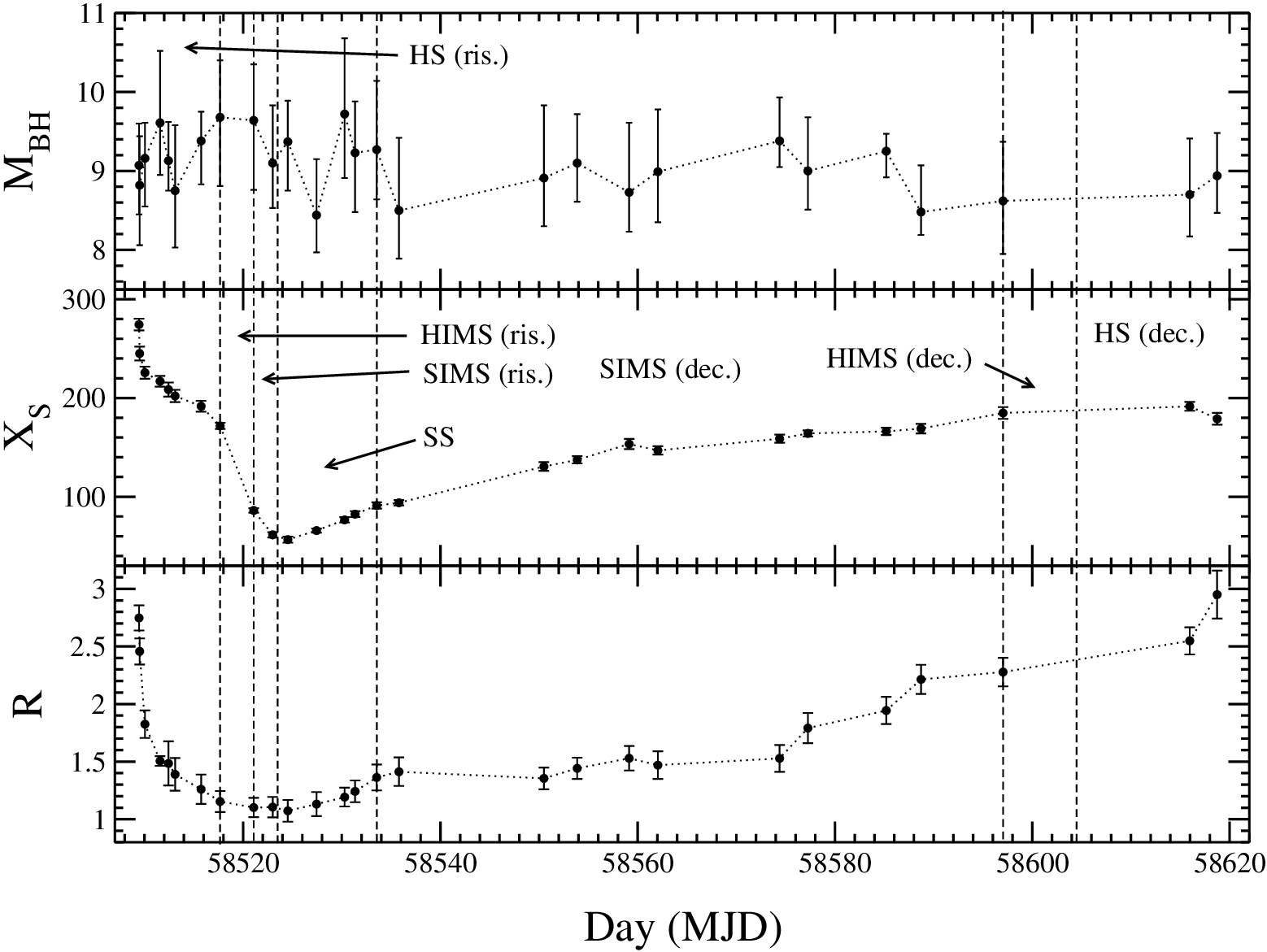}
\caption{The variation (a) the mass of the black hole ($M_{BH}$) in $M_{\odot}$, (b) the shock location 
($X_s$) in $r_s$, and (c) the shock compression ratio ($R$) are shown with time (MJD).
The verticle lines mark spectral state transition days.
\label{5}}
\end{figure*}

We used $1-150$ keV combined data from {\it Swift}/XRT, {\it Swift}/BAT and {\it MAXI}/GSC for the spectral analysis
of the BHC MAXI~J1348-630 during its 2019 outburst. We used the TCAF model-based fits file for the spectral study. 
Along with the TCAF model, we used $TBabs$ for absorption (Wilms et a. 2000) and $Gaussian$ for the line emission. 
A $Gaussian$ function at $\sim6.4$~keV was used to incorporate the $Fe-K{\alpha}$ emission line. In general, we used 
$TBabs(TCAF + gauss)$ model for the spectral analysis. We kept the hydrogen column density ($N_H$) free during our
analysis. We found it to vary between $0.58 \times 10^{22}$ cm$^{-2}$ and $3.24 \times 10^{22}$ cm$^{-2}$ during the
observation period. In Figure~\ref{spectra}, we show the TCAF model fitted spectra in $1-150$~keV energy range for
four different spectral states. Spectra in the hard state (HS) in the rising phase (2019 January 27, Obs. ID -  
00885960000), hard intermediate state (HIMS) in the rising phase (2019 February 03, Obs. ID - 00011107002), soft
state (SS) (2019 February 10, Obs. ID - 00011107007), and soft intermediate state (SIMS) in the declining phase (2019 
April 04, Obs. ID - 00896552000) are shown in panels - (a), (b), (c), and (d), respectively. The residuals obtained from 
the TCAF model fit are shown in the bottom panels of each spectrum. As quoted in the previous section, the TCAF model 
fit yields several parameters such as the Keplerian disk accretion rate ($\dot{m}_d$), the sub-Keplerian halo accretion 
rate ($\dot{m}_h$), the shock location ($X_s$), the shock compression ratio ($R$) etc. for each observation in the 
present work. The evolution of (a) the total accretion rate ($\dot{m}_d$+$\dot{m}_h$), (b) the Keplerian disk accretion
rate ($\dot{m}_d$), (c) the sub-Keplerian halo accretion rate ($\dot{m}_h$) and (d) the accretion rate ratio (ARR =
$\dot{m}_h$/$\dot{m}_d$) during the outburst are shown in Figure~\ref{4}(a-d). In Figure~\ref{5}, we show the variation
of the TCAF model fitted mass of the BH, the evolution of the shock location ($X_s$), and the shock compression ratio 
($R$) in panels-(a), (b) and (c), respectively. In Figure~\ref{mass}, we plot the variation of $\Delta \chi^2$ with 
the derived mass of the black hole ($M_{BH}$) for observations with Obs. IDs (a) 00885960000, (b) 00011107002, 
(c) 00011107007, and (d) 00896552000 taken from hard state, hard intermediate state, soft state and soft intermediate
state, respectively. The best-fitted parameters obtained from the TCAF model fitting to the source spectra during the 
outburst are presented in Table~1.

\subsection{Evolution of the Spectral State}
Earlier, we tried to identify the spectral states based on the variation of HRs, soft X-ray flux 
and hard X-ray flux. Here, we discuss the evolution of the spectral states based on the variation of 
the accretion rate ratio (ARR) and two types of accretion rates. Together with the evolution of HRs
and X-Ray fluxes, we classified the 2019 outburst of MAXI~J1348-630 in four usual spectral states:
HS, HIMS, SIMS, and SS. The detailed properties of the observed spectral states are mentioned in 
the following sub-Sections.

\subsubsection{Hard state in the rising phase -- HS (ris.)}
The source was in the HS when the observation started on 2019 January 26 (MJD 58509.45). The accretion
rates (both $\dot{m}_d$ and $\dot{m}_h$) increased in this state. High accretion rate ratio (ARR) was 
observed as $\dot{m}_h$ was higher than $\dot{m}_d$ in this state. As the day progressed, the ARR decreased 
slowly as $\dot{m}_d$ started to rise gradually. The shock was strong and moved rapidly in this state from
$274$ $r_s$ to $175$ $r_s$. We observed QPOs on 2019 January 29, and 2019 January 30. Higher HRs 
were also observed during this phase of the outburst as the high energy fluxes ($>4~keV$) in the GSC ($4-10$~keV), 
and BAT ($15-50$~keV) bands were dominated over the soft X-ray GSC band ($2-4$~keV). We marked 2019 February
3 (MJD 58517.67) as the transition day from the HS (ris.) to the HIMS (ris.) as after this day, $\dot{m}_d$
increased rapidly. We also found this day as the transition day from the evolution of HRs.

\subsubsection{Hard intermediate state in the rising phase -- HIMS (ris.)}

The source remained in this state till 2019 February 7 (MJD 58521.07). There are only two good {\it Swift}/XRT
observations of the source during this state. From the spectral fitting, we found that during these observations,
both the accretion rates ($\dot{m}_d$ and $\dot{m}_h$) increased and were comparable. The shock became weak with
decreasing $R$ and moved rapidly inward from $172$ $r_s$ to $86$ $r_s$. The ARR was roughly constant in this 
state, whereas the HR was found to decrease. The $\dot{m}_h$ became maximum on 2019 February 7 (MJD 58521.07)
when the source entered from the HIMS (ris.) to the SIMS (Dec.). 

\subsubsection{Soft intermediate state in the rising phase -- SIMS (ris.)}
 
During this spectral phase, the Keplerian disk accretion rate $\dot{m}_d$ increased rapidly. A decreasing trend
of the $\dot{m}_h$ is observed during this phase of the outburst. Due to this opposite trend of the two types of 
accretion rates, the ARR was found to decrease rapidly. A weak shock ($R \sim 1.1$) was found to move towards the 
BH (from $86$ $r_s$ to $56$ $r_s$). The HRs were observed to decrease slowly in this state.

\subsubsection{Soft state -- SS}

The source entered in the SS when the soft ($2-4$~keV), as well as the total ($2-10$~keV)
GSC X-ray fluxes, were maximum on 2019 February 9 (MJD 58523.5). However, on this day, 
no data was available for the spectral study. The spectral analysis result indicates that
the source was in the SS on 2019 February 10 (MJD 58524.53). Together with the evolution 
of ARR, HRs, GSC fluxes, and BAT flux, we conclude that the source entered in the SS on 
2019 February 9 (MJD 58523.5). In this state, high dominance of $\dot{m}_d$ over $\dot{m}_h$ 
was observed. This leads to cooling of the CENBOL and quenching of outflows if any. 
The Keplerian disk accretion rate reached its maximum value on 2019 February 10 (MJD 58524.53). 
After that, the Keplerian disk receded and the disk accretion rate decreased.
Rise of the shock parameters ($X_s$ and $R$) and ARR was observed on 2019 February 19 
(MJD 58533.54) when the source entered to the SIMS (dec.).

\subsubsection{Soft intermediate state in the declining phase -- SIMS (dec.)}

The source remained in the SIMS (dec.) till 2019 April 24 (MJD 58597.04). The shock was 
found to move outward (from $91$~$r_s$ to $185$~$r_s$). The shock also became strong ($R$ 
varied from $1.36$ to $2.55$) as the outburst progressed. Although both types of accretion
rates decreased with time, we saw an increasing trend in the ARR (from $0.41$ to $0.83$).
This is because  $\dot{m}_d$ decreased faster than the $\dot{m}_h$. This is also prominent
if we look at the variation of soft and hard fluxes and their ratios in Figure~\ref{profile}.

\subsubsection{Hard intermediate state in the declining phase -- HIMS (dec.)}

The source entered this spectral state on 2019 April 24 (MJD 58597.04). We observed a sharp rise in both HRs, 
as the hard X-ray ($4-10$~keV GSC and $15-50$~keV BAT) fluxes increased and soft X-ray ($2-4$~keV GSC) flux 
decreased (see Figure~\ref{profile}). No spectral data were available during this phase of the outburst.
The source entered the hard state in the declining phase on 2019 May 1 (MJD 58604.50). The transition 
day was marked based on the evolution of the HRs (see \S 3.2).

\subsubsection{Hard state in the declining phase -- HS (dec.)}

The source was in this state until the end of the observations. During this phase, HRs were roughly
constant. Both accretion rates ($\dot{m}_d$ \& $\dot{m}_h$) decreased monotonically in this state.
The ARR was found to increase. The shock became strong and moved away from the black hole.

\subsection{Viscous Time Scale}

Viscous time scale is the time at which high viscous matter reaches the BH from the pile-up radius
(Chakrabarti et al. 2019). In a transonic flow, a critical viscosity parameter ($\alpha_{crit}$) 
segregates two types of accretion flows. The high viscous Keplerian disk matter moves in viscous 
time scale along the equatorial plane, whereas low viscous, sub-Keplerian halo matter moves inward 
roughly in a free-fall time scale. As the halo matter moves faster than the disk, we observe that the halo
accretion rate attains its peak before the disk accretion rate. Differences in days when peaks occur
in these two types of flows give an estimation of the viscous time scale of the source (see, Jana et 
al. 2016). 

It is observed from Figure~4 that during the present outburst of MAXI~J1348-630, the halo accretion rate
became maximum on 2019 February 7 (MJD 58521.07) roughly $\sim 3.5$~days prior to that of the disk
accretion rate on 2019 February 10 (MJD 58524.53). This peak difference can be interpreted as the 
viscous timescale inside the Keplerian component.

\subsection{Estimation of the BH Mass}

Mass of the BH is a model input parameter in the TCAF model. If the mass is known, one can keep the mass of 
the BH frozen. Otherwise, it can be kept as a free parameter. The mass of MAXI~J1348-630 is not known 
and is allowed to vary while fitting with the TCAF model. Each observation gave us a best-fitted value 
of the mass. During the entire outburst, $M_{BH}$ best-fitted values varied between $8.44$~$M_{\odot}$ 
and $9.72$~$M_{\odot}$. Taking an average of these model fitted values, we obtained the average mass of 
MAXI~J1348-630 to be $9.1$~$M_{\odot}$.

\begin{figure}
\vskip 0.2cm
\centering
\includegraphics[width=9cm,keepaspectratio=true]{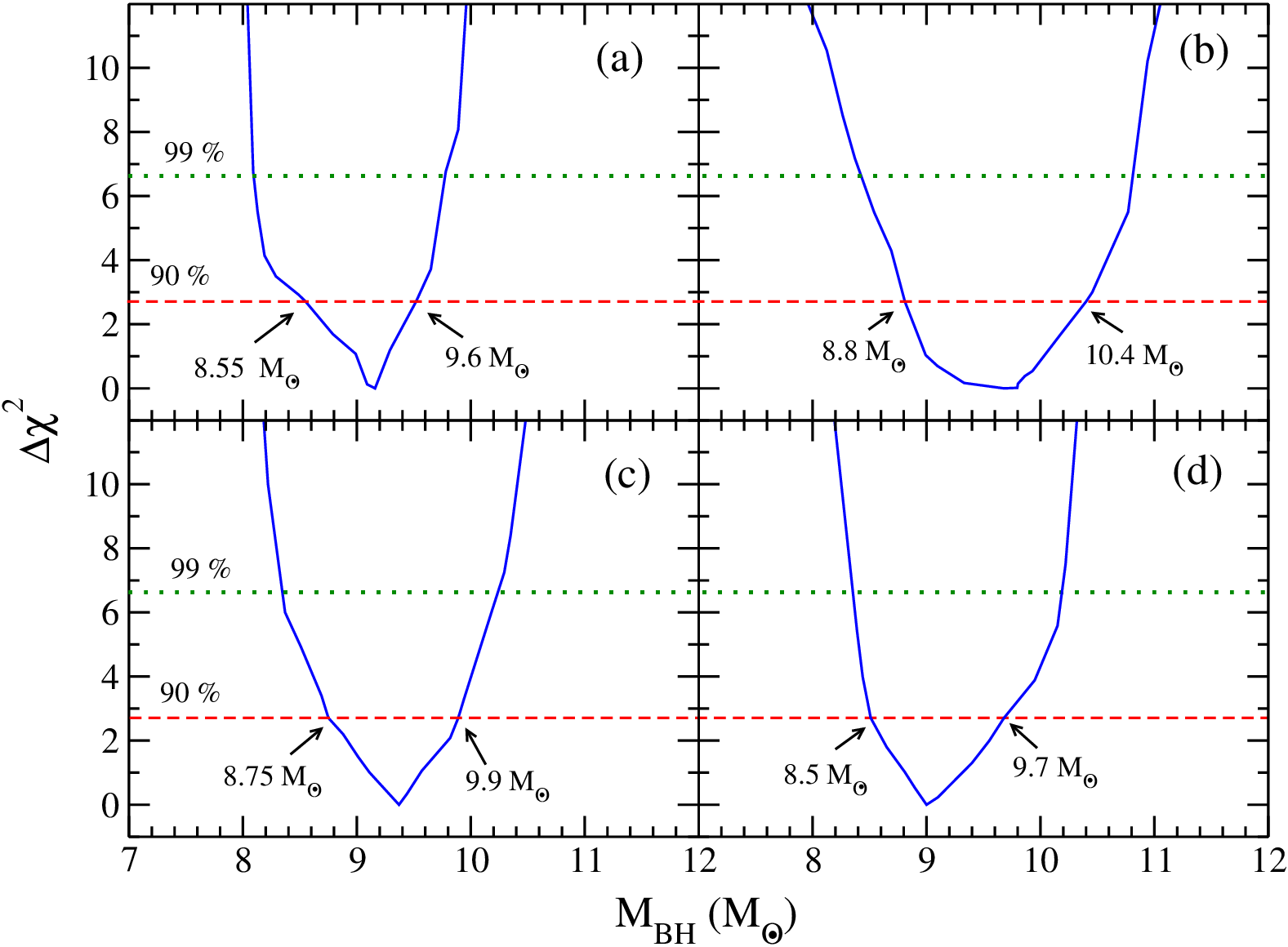}
\caption{The variation of $\Delta \chi^2$ with mass of the black hole ($M_{BH}$) in $M_{\odot}$ for observation
ID's (a) 00885960000, (b) 00011107002, (c) 00011107007, and (d) 00896552000 taken from hard state,
hard intermediate state, soft state and soft intermediate state respectively.
\label{mass}}
\end{figure}

We also used $M_{BH}-\Delta \chi^2$ method to estimate the mass of the BH (Molla et al. 2016, Chatterjee 
et al. 2016). In this process, we kept the mass frozen at different grid values on either side of the average
fitted value and checked how the $\Delta \chi^2$ varied with the changes of mass. In Figure~\ref{mass}, 
we showed these variations for four observations selected from four spectral states. With 90\% confidence 
($\Delta \chi^2 = 2.71$), we found the mass of the BH to be between $7.9$~$M_{\odot}$ and $10.7$~$M_{\odot}$. 
Combining these two methods, we estimated the probable mass of the BHC MAXI~J1348-630 to be 
$9.1^{+1.6}_{-1.2}$~$M_{\odot}$.

\subsection{Estimation of Distance}

We tried to estimate the distance of the source from the spectral state transition luminosity (Maccarone 2003; 
Tetarenko et al. 2016). Maccarone (2003) showed that the soft-to-hard state transition luminosity ($L_t$) is 
between $\sim$ 1\% and 4\% of Eddington luminosity ($L_{Edd}$). The Eddington luminosity is given by $1.3 
\times 10^{38}$ $(M/M_{\odot})$ ergs/s. We obtained the mass of MAXI J1348-630 as 9.1 $M_{\odot}$. So, Eddington 
luminosity of the source is $1.18 \times 10^{39}$ ergs/s. We considered MJD 58597.04 as the transition day 
between soft state to hard state as this day was marked as the transition day between SIMS (dec.) and HIMS (dec.). 
From $1-10$~keV XRT flux, we calculated the transition luminosity ($L_t$) of the source as $\sim 1.24 \times 10^{37}$ 
($d$/5 kpc) ergs/s, where $d$ is the distance of the source in kpc. For the source distance 5 kpc, 8 kpc, 10 kpc 
and 12 kpc, $L_t / L_{Edd}$ is obtained at 0.010, 0.026, 0.042 and 0.061 respectively. Therefore, we could predict 
the source distance as $5-10$ kpc.

\section{Discussion and Concluding Remarks}

We studied the evolution of the timing and the spectral properties of newly discovered Galactic transient
BHC MAXI~J1348-630, during its 2019 outburst. We studied the BHC for a duration of about 4 months using data 
from {\it Swift}/XRT, {\it Swift}/BAT and {\it MAXI}/GSC instruments. For the spectral analysis, we used the 
TCAF model-based fits file. From the TCAF model, we extracted flow parameters of the system such as the Keplerian 
disk accretion rate ($\dot{m}_d$), the sub-Keplerian halo accretion rate ($\dot{m}_h$), the shock location ($X_s$)
and shock compression ratio ($R$). We also estimated the mass of the BH from our spectral analysis. These 
parameters were obtained from each observation, collectively giving us an idea of accretion flow dynamics 
during the observational period.

It is to be noted that we used XRT ($1-10$~keV), GSC ($7-20$~keV), and BAT ($15-150$~keV) data to cover 
broad energy range of $1-150$~keV. It is best to use a broad energy range to obtain the flow parameters
more accurately with TCAF. In general, without BAT, $1-10$~keV XRT data or $1-20$~keV XRT+GSC data were 
available for the spectral analysis. Therefore, part of the spectral information was missing when BAT 
data were absent. This is the case for any model and not just TCAF. However, the TCAF model handles both
narrow and broadband data better than any phenomenological model. TCAF calculates all the model parameters
self-consistently. Therefore, we do not expect the fitting result to be changed abruptly with or without BAT.
We checked it by removing BAT data from broadband spectra (when XRT+BAT was used). We found that the results
did not change significantly without BAT. As an example, on the first observation (Obs ID: 00885807000),
we obtained $\dot{m_d} = 0.36$ , $\dot{m_h} = 0.41$, $X_s = 274$, $R = 3.07$ for $\chi^2$/dof=312/287 with
$1-150$ keV XRT+GSC+BAT data. When we removed BAT data, we obtained $\dot{m_d} = 0.36$ , $\dot{m_h} = 0.41$,
$X_s = 272$, $R = 3.06$ for $\chi^2$/dof=254/231 with $1-20$ keV XRT+GSC data. The fit result and statistics
did not change significantly when we did not use BAT data.

MAXI~J1348-630 went into its first outburst in 2019. The source was discovered on 2019 January 26 
(MJD 58509.45). The outburst continued for $\sim 4$ months. Both the soft and the hard X-ray intensities
increased in the rising phase of the outburst, although at different rates. Hard X-ray flux ($15-50$~keV 
BAT) showed a rapid rise and reached its peak flux on 2019 February 3 (MJD 58517.67). After that, BAT flux
declined. The soft X-ray ($2-4$~keV GSC) flux increased monotonically and achieved peak value on 2019 February
9 (MJD 58523.5). On that day, the BHC entered into the SS. The intermediate period (from MJD 58517.67 to 
MJD 58523.5) belongs to two intermediate spectral states (rising HIMS and SIMS). In the declining phase,
the X-ray flux slowly decreased. Depending upon the rising (slow) and decreasing (slow) variation of the 
total X-ray flux ($2-10$~keV GSC), it appears that the nature of the outburst is a `slow-rise-slow-decay' 
(SRSD) type (see, Debnath et al. 2010). 

We identified the spectral states into two independent ways: (i) from the evolution of the HRs, the soft 
X-ray flux and the hard X-ray flux, and (ii) from the variation of the TCAF model fitted spectral parameters. 
However, to mark the exact state transition days, we used both methods. For example, the source entered in 
the SS from the SIMS (ris.) on MJD 58523.5. However, no spectral data were available on that day. Spectral 
data were available on MJD 58524.53, and the source was clearly in the SS on that day. Similarly, the 
transition day between HIMS (dec.) to HS (dec.) was marked from the evolution of HRs. During the outburst, 
MAXI~J1348-630 showed all four usual spectral states in the following sequence: HS (ris.) $\rightarrow$ HIMS 
(ris.) $\rightarrow$ SIMS (ris.)$ \rightarrow$ SS $\rightarrow$ SIMS (dec.) $\rightarrow$ HIMS (dec.) 
$\rightarrow$ HS (dec.).  

MAXI~J1348-630 was in the HS when the outburst started. The sub-Keplerian halo accretion rate was higher 
than the Keplerian disk accretion rate in this state. A strong shock was found at a significant distance 
from the BH. Both accretion rates increased as the outburst progressed. We see the peak of the halo accretion 
rate on HIMS (ris.) to SIMS (ris.) transition day (2019 February 7; MJD 58521.07). The Keplerian disk accretion 
rate increased rapidly than the sub-Keplerian halo accretion rate from this transition day. More supply in the 
disk matter cooled the CENBOL in a faster way as the source moved to the SIMS (ris.) and SS. In the SS, the 
total accretion rate achieved its peak value on 2019 February 10 (MJD 58524.53). Then the source entered into 
the declining phase. Both accretion rates decreased slowly. The source remained in the softer states (SIMS and
SS) for a long time ($\sim 2.5$ months). We have also noticed that in the present outburst of MAXI~J1348-630, 
the duration of the SIMS (dec.) was much longer compared to the SS and other spectral states. Generally, we see
longer SS in the outbursting BHCs. As the outburst progressed, the shock became strong and receded from the
BH starting from the SS to SIMS (dec.) transition day (2019 February 19; MJD 58533.54). The ARR also increased
slowly. This trend continued in the declining HIMS and HS too. In general, the ARR increased in the HS of the
declining phase as the spectra became hard.

Unlike many outbursting candidates, we did not observe QPO on each day of observation during 2019 
outburst of MAXI~J1348-630. We detected low-frequency QPOs only in two observations in the rising phase of 
the outburst. The QPOs were detected on 2019 January 29 and 30 at frequencies of $0.57\pm0.04$ Hz and $0.66\pm 0.04$ 
Hz with Q-values of $4.5\pm0.4$ and $2.8\pm0.3$, respectively. Chen et al. (2019) also found an evolving 
QPO on 2019 January 30 with maximum centroid frequency of $0.71 \pm 0.01$ Hz with FWHM of $0.16 \pm 0.04$ Hz 
from HXMT observation. They studied roughly for the full-day observation. Thus, we feel that our findings are
consistent with their reports. No QPO was observed in the declining phase of the outburst. The shock oscillations 
are believed to be the reason behind these QPOs. Since type-C low-frequency QPOs occur due to resonance between 
cooling and infall time scales (Molteni et al. 1996), rapid deterioration of Q-value indicates that the flow went 
off the resonance very quickly. Type-A or B QPOs are observed sporadically in the SIMS due to a weakly oscillating 
shock. This weak shock oscillation could be either due to a weakly resonating CENBOL (for type-B) or due to 
a shock-less centrifugal barrier (for type-A) (Ryu et al. 1997). In the present system, such QPOs were not observed.

We checked if the resonance condition is satisfied during the outburst. The resonance condition is satisfied
when the infall (compressional heating) timescale ($t_{in}$) of the post-shock matter matches with the cooling
timescale ($t_{c}$) of the post-shock matter (Molteni et al. 1996, Chakrabarti et al. 2015, hereafter CMD15). 
As far as the resonance condition is concerned, the ratio has to be strictly one to have resonance by definition. 
However, since the CENBOL is not oscillating as a single entity; rather, different parts are oscillating with
slightly different frequency, the resonance condition is achieved when the ratio is in the order of unity.
CMD15 showed that if the ratio is between 0.5 and 1.5, i.e., within 50\% of unity, the resonance is satisfied.
However, this limit is not rigid. Even outside of this range, some oscillation may occur, and instead of Type C 
QPOs, one may also see Type B or A QPOs. As the ratio deviates, chances of finding QPO with a significant decrease
of rms reduces, what we saw here. Jana et al. (2020b) applied this method to investigate the reason behind the 
non-observation of LFQPOs during the 2015 outburst of V404~Cygni. For the present object too, we followed the 
CMD15 procedure to calculate $t_c$ and $t_{in}$. We find that the resonance condition was satisfied 
only during three observations, i.e., on 2019 January 29 (MJD 58512.43), 2019 January 30 (MJD 58513.11) and 2019
February 1 (MJD = 58515.75) with the ratio being 0.94 and 0.79 and 0.60, respectively. The cooling times 
($t_c$) for three observations are 1.20 s, 0.96 s, and 0.67 s, respectively. The infall times ($t_{in}$) are 1.28 s, 
1.22 s and 1.12 s for MJD 58512.43, MJD 58513.11 and MJD 58515.75, respectively. The QPOs were indeed observed only 
in two observations, on MJD 58512.43 and MJD 58513.11. On MJD 58515.75, non-observation of the QPO could be due
to weak satisfaction of the resonance condition. In other observations, QPOs were not observed as the resonance 
condition was not satisfied due to efficient cooling processes. Such observation is consistent with the expectation 
from a purely theoretical point of view. 

Low viscous sub-Keplerian flow has low angular momentum, whereas the Keplerian disk has high angular momentum
and viscosity. Thus, the sub-Keplerian halo moves faster than the Keplerian disk, which moves in the viscous
timescale. Thus the sub-Keplerian flow rate achieves its peak earlier than the Keplerian disk rate. The time 
difference of reaching the maxima of the sub-Keplerian halo accretion rate and the Keplerian disk accretion 
rate is the viscous timescale of the Keplerian component. Jana et al. (2016) calculated the viscous timescale
for MAXI~J1836-194 during its 2011 outburst as $\sim 10$ days using this method. In the same way, viscous 
timescales are also calculated for a few other BHCs (Mondal et al. 2017). During the 2019 outburst of MAXI~J1348-630, 
the sub-Keplerian halo accretion rate became maximum on 2019 February 7 (MJD 58521.07), and the Keplerian 
disk accretion rate became maximum on 2019 February 10 (MJD 58524.53). This suggests that the viscous timescale
for the outburst is $\sim 3.5$ days.

During the spectral analysis, we found that the hydrogen column density ($N_H$) varied between $0.58\times 10^{22}$ 
cm$^{-2}$ and $3.24\times10^{22}$ cm$^{-2}$ during the outburst. It is observed that $N_H$ may change 
in different spectral states due to the radio activity and the outflowing matter from the disk (Sreehari
et al. 2019). If the accretion rate is high, there could be outflow from the disk due to radiation pressure
(Radhika et al. 2016a). This disk may lead to the variable $N_H$ (Radhika et al. 2016ab). In MAXI~J1348-630, 
the super-Eddington accretion rate was observed, which could lead to disk wind, hence variable $N_H$.

We observed Fe emission line in 11 observations out of a total 27 observations in our analysis.
The Fe line was detected in the rising phases of HS, HIMS \& SIMS, and in the SS. In the declining
phase, no Fe line was detected except for one observation. In general, the Fe line was detected 
when the Keplerian disk rate was high. This suggests that the Fe line may be associated with the
Keplerian disk. This could be the reason for non-detection of Fe line in the declining phase 
when the Keplerian disk receded and the disk rate decreased. In our observation, the line energy 
varied between $6.06$~keV and $6.95$~keV. The observed line around $\sim 6.9$~keV might be Fe~XXVI
line. The observed broad lines could be a possible blend of several Fe lines, which could not be 
resolved individually.

Tominaga et al. (2020) estimated the mass of the BH from the result obtained from the tbabs*simpl*diskbb
model. They calculated $R_{in}$ from the diskbb normalization and equated with ISCO. From this, they
calculated the mass of the BH to be $>16~M_{\odot}$. However, this method is not reliable, since the
disk would extend up to ISCO or 3 $r_s$ only in an ideal case, but often that is not achieved.
Phenomenological model fitting gives a rough idea of the system parameters, but a physical model is
required to study the dynamics of the system. Unlike the diskbb model, TCAF is a physical model, 
and all spectral parameters (including mass) are calculated self-consistently. In the TCAF model 
fits file, the mass of the BH is an important model input parameter, simply because of the electron 
number density in CENBOL, soft photon intensity from the Keplerian disk, the size of the CENBOL, etc.,
which are responsible for producing the spectrum, depend on the mass. Thus, it is possible to estimate
the mass of the BH from the spectral analysis with the physical TCAF model. The mass of several Galactic 
BHCs and AGN have already been estimated successfully from the spectral analysis with the TCAF model
(Molla et al. 2016, Chatterjee et al. 2016, 2019; Jana et al. 2016; Bhattacharjee et al. 2017;
Nandi et al., 2019). Jana et al. (2019) reported the mass of this BHC in the range of $8.5-11$~$M_{\odot}$ 
from their preliminary analysis. In this paper, we estimated the mass of the BHC MAXI~J1348-630 by 
keeping $M_{BH}$ free while fitting the spectra with the TCAF model. Each observation gave us a 
best-fitted $M_{BH}$ which was found to vary between $8.44$ $M_{\odot}$ and $9.72$ $M_{\odot}$ (see Appendix).
This variation is the result of poor data quality and errors in fitting the data. For instance, 
fitting the peak of the multicolor black body depends on disk temperature $T$, and errors introduced 
are amplified four times, since $M_{BH} \sim T^{-4}$. This is mainly contributing to the $\sim 4$ 
times error in $M_{BH}$. From the variation of the model fitted $M_{BH}$, we calculated the average 
value of the mass as $9.1$ $M_{\odot}$. We also checked the mass from $\Delta \chi^2$-$M_{BH}$ 
plots. We kept mass frozen at different values and checked how $\Delta \chi^2$ varied. From this,
we find that the mass of the BHC MAXI J1348-630 is between $7.9$ and $10.7$~$M_{\odot}$ with 
$90 \%$ confidence. Combining these two methods, we conclude the mass of the BHC MAXI~J1348-630 
to be in the range of $7.9-10.7$~$M_{\odot}$ or $9.1 ^{+1.6}_{-1.2}$~$M_{\odot}$.

We estimated the distance of the source from the hard-to-soft state transition luminosity. It is observed that 
the soft-to-hard transition luminosity ($L_t$) is between 0.01 $L_{Edd}$ and 0.04 $L_{Edd}$ (Maccarone 2003, 
Teterenko et al 2016). From this method, the distance of MAXI~J1910-057 was calculated to be $>1.70$ kpc 
(Nakahira et al. 2014). We calculated TCAF model fitted $1-10$~keV luminosity as $L = 1.24 \times 10^{37}$ 
(d/5) ergs/s. We found that for $d=5-10$, $L_t/L_{Edd} = 0.01-0.042$. This infers that the source distance is 
between $5-10$ kpc. In another emperical relation, McClintock \& Remillard (2009) showed that peak luminosity 
would be $0.2-0.4 L_{Edd}$ for three BHCs GRS~1915+105, GRO~J1655-40 and XTE~J1550-564. The peak luminosity of 
MAXI~J1348-630 was observed on MJD 58524.53 with $L_{peak} = 1.90\times 10^{39}$ (d/5) ergs/s. From this, 
$L_{peak}/L_{Edd} =$ 0.16, 0.23 and 0.41 for d = 5 kpc, 6 kpc and 8 kpc respectively. From this, the source 
distance is $5-8$ kpc. Tominaga et al. (2020) estimated the source distance as $4-8$ kpc using the same methods 
mentioned above. Our finding is consistent with the findings of Tominaga et al. (2020).

In the TCAF model, normalization $N$ remains constant across the spectral states simply because it is just
a scaling factor to convert the emitted spectrum to the TCAF spectrum (Molla et al. 2016; 2017). However, 
if a jet is present, normalization could vary, and $N$ is observed to have a higher value (Jana et al. 2017; 
2020a). In this case, emitted X-rays contain the contribution from the inner jet, which was theorized in TCAF.
During the 2019 outburst of BHC MAXI~J1348-630, normalization was not found to be a constant. This indicates
the presence of the X-ray jets during the outburst. The disk-jet connection of the source will be studied and
published elsewhere.

\section{Summary}
We studied MAXI~J1348-630 during its 2019 outburst in detail. We used data of {\it Swift}/XRT, {\it Swift}/BAT
and {\it MAXI}/GSC in the combined broad energy range of $1-150$~keV for our study. We find that the source went 
through all the usual spectral states. We presented how the flow parameters evolved during the outburst from TCAF
fits of the spectra. We observed QPOs only in two observations when the stricter resonance condition between the
heating and the cooling times scales of the Compton cloud were found to be satisfied. We estimated the mass of 
MAXI~J1348-630 is $9.1^{+1.6}_{-1.2}$~$M_{\odot}$. We also estimated the viscous timescale of the standard disk 
component to be $\sim 3.5$ days during the outburst. From the state transition luminosity, we estimated the 
distance of the source as $5-10$ kpc.

\section*{Acknowledgment}
We thank anonymous referee for his/her comments and suggestions to improve the quality of this manuscript.
This work was made use of XRT and BAT data supplied by the UK Swift Science Data Centre at the University of Leicester, 
and MAXI data was provided by RIKEN, JAXA, and the MAXI team.
A.J. acknowledges post-doctoral fellowship of Physical Research Laboratory, Ahmedabad, funded by the Department of Space, 
Government of India.
D.D. and A.J. acknowledge support from DST/GITA sponsored India-Taiwan collaborative project (GITA/DST/TWN/P-76/2017) fund.
Research of D.D. and S.K.C. is supported in part by the Higher Education Dept. of the Govt. of West Bengal, India.
S.K.C. and D.D. also acknowledge partial support from ISRO sponsored RESPOND project (ISRO/RES/2/418/17-18) fund.
K.C. acknowledges support from DST/INSPIRE Fellowship (IF170233).
R.B. acknowledges support from CSIR-UGC NET qualified UGC fellowship (June-2018, 527223).
N. K. acknowledges support from the research fellowship from Physical Research Laboratory, Ahmedabad, India, funded by 
the Department of Space, Government of India.

\clearpage

\begin{longrotatetable}
\vskip -1.5cm
\addtolength{\tabcolsep}{-5.50pt}
\small
\centering
\centering{\large \bf Table I}
\vskip -0.6cm
\centerline {}
\vskip 0.0cm
\begin{tabular}{lcccccccccccccccc}
\hline
Obs ID& UT Date&Day&XRT exp$^a$&~~BAT exp$^a$&$N_H$~$^b$&$\dot{m_d}^c$&$\dot{m_h}^c$&ARR&$X_{S}$ $^d$& $R$ &$N$&LE$^e$&LW$^e$&LN&$\chi^2 /dof$\\
&(mm-dd) & (MJD) & (sec)& (sec) &$(10^{22}$ cm$^{-2})$ &  ($\dot{M}_{Edd}$) & ($\dot{M}_{Edd}$) & & ($r_s$) & & & (keV) & (keV)&(ph/cm$^2$/s) & \\
(1)&  (2)  & (3)  & (4)& (5) & (6) & (7)&(8) &(9) &(10) &(11) &(12) &(13) &(14)&(15)&(16) \\
\hline
00885807000& 01-26 & 58509.45 & 1223$^*$& 3794 &~ $1.09^{+0.14}_{-0.17} $&~ $ 0.36^{+0.02}_{-0.02} $&~ $ 0.41^{+0.03}_{-0.02} $&~$ 1.15 ^{\pm0.12}$ &~ $274^{+7}_{-3}$&~ $ 3.07^{+0.12}_{-0.13}$&~$ 268 ^{+23}_{-25} $&~ $   -                  $&~ $ -                  $&$      -             $& 312/287 \\
00885845000& 01-26 & 58509.51 & --      & 2945 &~ $0.88^{+0.31}_{-0.22} $&~ $ 0.45^{+0.01}_{-0.03} $&~ $ 0.52^{+0.02}_{-0.02} $&~$ 1.17 ^{\pm0.08}$ &~ $245^{+5}_{-9}$&~ $ 2.46^{+0.11}_{-0.12}$&~$ 401 ^{+28}_{-23} $&~ $   -                  $&~ $ -                  $&$      -             $& 57 /51  \\
00885960000& 01-27 & 58510.05 & 1455$^*$& 4907 &~ $0.66^{+0.11}_{-0.9 } $&~ $ 0.54^{+0.04}_{-0.05} $&~ $ 0.55^{+0.03}_{-0.01} $&~$ 1.02 ^{\pm0.12}$ &~ $225^{+4}_{-8}$&~ $ 1.82^{+0.14}_{-0.10}$&~$ 295 ^{+15}_{-22} $&~ $   -                  $&~ $ -                  $&$      -             $& 314/265 \\
00886266000& 01-28 & 58511.58 & 1838$^*$& 3838 &~ $0.63^{+0.24}_{-0.21} $&~ $ 0.55^{+0.05}_{-0.05} $&~ $ 0.56^{+0.04}_{-0.02} $&~$ 1.02 ^{\pm0.11}$ &~ $217^{+5}_{-5}$&~ $ 1.50^{+0.04}_{-0.03}$&~$ 202 ^{+22}_{-20} $&~ $ 6.16 ^{+0.14}_{-0.05}$&~ $1.93^{+0.05}_{-0.06}$&$0.31^{+0.04}_{-0.03}$& 302/262 \\
00886496000& 01-29 & 58512.43 &  317$^*$& 2781 &~ $0.97^{+0.13}_{-0.14} $&~ $ 0.62^{+0.05}_{-0.07} $&~ $ 0.58^{+0.03}_{-0.01} $&~$ 0.96 ^{\pm0.13}$ &~ $209^{+6}_{-7}$&~ $ 1.48^{+0.13}_{-0.16}$&~$ 341 ^{+35}_{-39} $&~ $   -                  $&~ $ -                  $&$       -            $& 328/266 \\
00011107001& 01-30 & 58513.11 &  834$^*$& --   &~ $1.05^{+0.12}_{-0.10} $&~ $ 0.66^{+0.03}_{-0.04} $&~ $ 0.59^{+0.04}_{-0.02} $&~$ 0.90 ^{\pm0.08}$ &~ $202^{+5}_{-6}$&~ $ 1.39^{+0.10}_{-0.15}$&~$ 159 ^{+15}_{-21} $&~ $ 6.59 ^{+0.13}_{-0.11}$&~ $1.00^{+0.08}_{-0.04}$&$0.29^{+0.04}_{-0.05}$& 728/583 \\
00088843001& 02-01 & 58515.75 & 2033$^*$& --   &~ $1.16^{+0.24}_{-0.21} $&~ $ 0.71^{+0.06}_{-0.07} $&~ $ 0.59^{+0.02}_{-0.03} $&~$ 0.83 ^{\pm0.09}$ &~ $191^{+4}_{-7}$&~ $ 1.24^{+0.10}_{-0.04}$&~$ 870 ^{+53}_{-37} $&~ $ 6.74 ^{+0.14}_{-0.10}$&~ $1.93^{+0.10}_{-0.07}$&$0.28^{+0.03}_{-0.04}$& 357/297 \\
\hline
00011107002& 02-03 & 58517.67 &  870$^*$& --   &~ $0.58^{+0.11}_{-0.10} $&~ $ 0.76^{+0.09}_{-0.06} $&~ $ 0.62^{+0.03}_{-0.04} $&~$ 0.82 ^{\pm0.09}$ &~ $172^{+5}_{-3}$&~ $ 1.15^{+0.10}_{-0.05}$&~$ 288 ^{+33}_{-44} $&~ $ 6.16 ^{+0.15}_{-0.07}$&~ $1.00^{+0.12}_{-0.09}$&$0.15^{+0.03}_{-0.02}$& 229/181 \\
\hline
00011107004& 02-07 & 58521.07 & 1749$^*$& --   &~ $1.49^{+0.22}_{-0.25} $&~ $ 0.95^{+0.08}_{-0.09} $&~ $ 0.73^{+0.04}_{-0.02} $&~$ 0.77 ^{\pm0.08}$ &~ $86 ^{+2}_{-2}$&~ $ 1.10^{+0.08}_{-0.05}$&~$ 134 ^{+22}_{-25} $&~ $ 6.90 ^{+0.07}_{-0.13}$&~ $1.00^{+0.10}_{-0.07}$&$0.10^{+0.02}_{-0.01}$& 142/128 \\
00011107005& 02-09 & 58523.00 & 1414$^*$& --   &~ $1.33^{+0.23}_{-0.18} $&~ $ 1.34^{+0.08}_{-0.08} $&~ $ 0.71^{+0.05}_{-0.03} $&~$ 0.53 ^{\pm0.05}$ &~ $61 ^{+3}_{-5}$&~ $ 1.10^{+0.07}_{-0.03}$&~$ 164 ^{+16}_{-24} $&~ $ 6.93 ^{+0.10}_{-0.15}$&~ $0.83^{+0.07}_{-0.07}$&$0.12^{+0.02}_{-0.01}$& 129/130 \\
\hline
00011107007& 02-10 & 58524.53 &  809$^*$& --   &~ $1.88^{+0.34}_{-0.47} $&~ $ 1.64^{+0.10}_{-0.09} $&~ $ 0.62^{+0.04}_{-0.03} $&~$ 0.38 ^{\pm0.04}$ &~ $56 ^{+2}_{-4}$&~ $ 1.09^{+0.07}_{-0.04}$&~$ 323 ^{+40}_{-54} $&~ $ 6.75 ^{+0.22}_{-0.25}$&~ $0.71^{+0.03}_{-0.05}$&$0.11^{+0.01}_{-0.03}$& 238/209 \\
00011107010& 02-13 & 58527.44 & 1465$^*$& --   &~ $2.43^{+0.27}_{-0.37} $&~ $ 1.55^{+0.11}_{-0.12} $&~ $ 0.50^{+0.03}_{-0.03} $&~$ 0.32 ^{\pm0.03}$ &~ $66 ^{+4}_{-3}$&~ $ 1.13^{+0.10}_{-0.34}$&~$ 55  ^{+18}_{-12} $&~ $ 6.93 ^{+0.17}_{-0.19}$&~ $0.88^{+0.10}_{-0.09}$&$0.15^{+0.02}_{-0.03}$& 122/133 \\
00011107011& 02-16 & 58530.29 & 1464$^*$& --   &~ $1.20^{+0.23}_{-0.27} $&~ $ 1.44^{+0.08}_{-0.11} $&~ $ 0.44^{+0.04}_{-0.01} $&~$ 0.31 ^{\pm0.02}$ &~ $77 ^{+2}_{-3}$&~ $ 1.19^{+0.08}_{-0.24}$&~$ 85  ^{+12}_{-11} $&~ $ 6.67 ^{+0.18}_{-0.21}$&~ $0.64^{+0.05}_{-0.05}$&$0.19^{+0.02}_{-0.03}$& 141/136 \\
00011107012& 02-17 & 58531.35 & 1460$^*$& --   &~ $0.89^{+0.19}_{-0.17} $&~ $ 1.29^{+0.12}_{-0.11} $&~ $ 0.45^{+0.02}_{-0.04} $&~$ 0.34 ^{\pm0.04}$ &~ $82 ^{+4}_{-5}$&~ $ 1.24^{+0.09}_{-0.15}$&~$ 41  ^{+15}_{-13} $&~ $ 6.42 ^{+0.13}_{-0.17}$&~ $0.71^{+0.05}_{-0.06}$&$0.18^{+0.02}_{-0.01}$& 130/136 \\
\hline
00011107013& 02-19 & 58533.54 & 1024$^*$& --   &~ $3.24^{+0.30}_{-0.41} $&~ $ 1.13^{+0.08}_{-0.07} $&~ $ 0.47^{+0.03}_{-0.04} $&~$ 0.41 ^{\pm0.04}$ &~ $91 ^{+4}_{-5}$&~ $ 1.36^{+0.12}_{-0.13}$&~$ 59  ^{+15}_{-12} $&~ $ -                    $&~ $ -                  $&$    -               $& 255/224 \\
00011107014& 02-21 & 58535.80 & 1129$^*$& --   &~ $1.26^{+0.21}_{-0.25} $&~ $ 1.09^{+0.10}_{-0.12} $&~ $ 0.41^{+0.02}_{-0.02} $&~$ 0.37 ^{\pm0.04}$ &~ $94 ^{+3}_{-3}$&~ $ 1.41^{+0.11}_{-0.14}$&~$ 117 ^{+15}_{-11} $&~ $ -                    $&~ $ -                  $&$    -               $& 269/224 \\ 
00088843002& 03-08 & 58550.49 & 2079    & --   &~ $1.79^{+0.39}_{-0.36} $&~ $ 0.77^{+0.07}_{-0.08} $&~ $ 0.33^{+0.03}_{-0.02} $&~$ 0.43 ^{\pm0.04}$ &~ $131^{+3}_{-4}$&~ $ 1.35^{+0.09}_{-0.09}$&~$ 301 ^{+37}_{-50} $&~ $ -                    $&~ $ -                  $&$    -               $& 271/212 \\ 
00011107017& 03-11 & 58553.87 &  955    & --   &~ $2.77^{+0.41}_{-0.46} $&~ $ 0.64^{+0.03}_{-0.06} $&~ $ 0.32^{+0.02}_{-0.01} $&~$ 0.50 ^{\pm0.06}$ &~ $137^{+2}_{-6}$&~ $ 1.44^{+0.09}_{-0.10}$&~$ 476 ^{+40}_{-43} $&~ $ -                    $&~ $ -                  $&$    -               $& 252/218 \\ 
00011107019& 03-17 & 58559.13 & 1188    & --   &~ $2.07^{+0.32}_{-0.40} $&~ $ 0.52^{+0.03}_{-0.03} $&~ $ 0.27^{+0.02}_{-0.03} $&~$ 0.52 ^{\pm0.06}$ &~ $153^{+4}_{-6}$&~ $ 1.53^{+0.11}_{-0.11}$&~$ 801 ^{+44}_{-58} $&~ $ -                    $&~ $ -                  $&$    -               $& 961/891 \\ 
00011107020& 03-20 & 58562.04 & 1001    & --   &~ $1.61^{+0.22}_{-0.29} $&~ $ 0.39^{+0.04}_{-0.03} $&~ $ 0.21^{+0.02}_{-0.02} $&~$ 0.54 ^{\pm0.07}$ &~ $147^{+6}_{-4}$&~ $ 1.47^{+0.12}_{-0.08}$&~$ 312 ^{+39}_{-27} $&~ $ 6.06 ^{+0.13}_{-0.10}$&~ $0.49^{+0.04}_{-0.04}$&$0.08^{+0.02}_{-0.01}$&1021/880 \\
00011107024& 04-01 & 58574.38 & 1025    & --   &~ $2.66^{+0.39}_{-0.30} $&~ $ 0.30^{+0.02}_{-0.03} $&~ $ 0.17^{+0.01}_{-0.01} $&~$ 0.57 ^{\pm0.07}$ &~ $159^{+7}_{-5}$&~ $ 1.53^{+0.13}_{-0.10}$&~$ 96  ^{+16}_{-13} $&~ $ -                    $&~ $-                   $&$    -               $& 904/885 \\
00896552000& 04-04 & 58577.25 & 1575    & 3893 &~ $1.82^{+0.22}_{-0.27} $&~ $ 0.26^{+0.01}_{-0.01} $&~ $ 0.17^{+0.01}_{-0.02} $&~$ 0.66 ^{\pm0.07}$ &~ $163^{+6}_{-8}$&~ $ 1.79^{+0.11}_{-0.15}$&~$ 294 ^{+35}_{-27} $&~ $ -                    $&~ $-                   $&$                    $& 285/267 \\
00011107027& 04-12 & 58585.20 &  980    & --   &~ $2.32^{+0.26}_{-0.31} $&~ $ 0.15^{+0.01}_{-0.02} $&~ $ 0.12^{+0.01}_{-0.01} $&~$ 0.81 ^{\pm0.10}$ &~ $164^{+5}_{-2}$&~ $ 2.02^{+0.12}_{-0.14}$&~$ 401 ^{+38}_{-31} $&~ $ -                    $&~ $-                   $&$    -               $&1174/892 \\
00011107028& 04-15 & 58588.72 & 1020    & --   &~ $1.47^{+0.15}_{-0.18} $&~ $ 0.14^{+0.01}_{-0.01} $&~ $ 0.11^{+0.01}_{-0.01} $&~$ 0.83 ^{\pm0.11}$ &~ $166^{+6}_{-7}$&~ $ 2.25^{+0.10}_{-0.13}$&~$ 213 ^{+16}_{-18} $&~ $ -                    $&~ $-                   $&$    -               $& 995/892 \\ 
\hline
00011107029& 04-24 & 58597.04 & 1015    & --   &~ $1.63^{+0.24}_{-0.18} $&~ $ 0.13^{+0.01}_{-0.01} $&~ $ 0.11^{+0.00}_{-0.01} $&~$ 0.85 ^{\pm0.10}$ &~ $185^{+8}_{-6}$&~ $ 2.28^{+0.13}_{-0.11}$&~$ 287 ^{+42}_{-35} $&~ $ -                    $&~ $-                   $&$    -               $& 969/899 \\
\hline
00011107035& 05-12 & 58615.97 &  604    & --   &~ $1.09^{+0.14}_{-0.17} $&~ $ 0.08^{+0.01}_{-0.01} $&~ $ 0.09^{+0.01}_{-0.01} $&~$ 1.01 ^{\pm0.14}$ &~ $192^{+5}_{-7}$&~ $ 2.55^{+0.14}_{-0.10}$&~$ 116 ^{+13}_{-16} $&~ $ -                    $&~ $-                   $&$    -               $& 682/917 \\
00011107036& 05-15 & 58618.74 & 1079    & --   &~ $1.37^{+0.20}_{-0.22} $&~ $ 0.07^{+0.01}_{-0.01} $&~ $ 0.08^{+0.01}_{-0.01} $&~$ 1.10 ^{\pm0.19}$ &~ $179^{+7}_{-6}$&~ $ 2.95^{+0.15}_{-0.17}$&~$ 277 ^{+31}_{-37} $&~ $ -                    $&~ $-                   $&$    -               $& 698/892 \\ 

\hline
\end{tabular}
\noindent{
\leftline{`$^*$' indicates {\it MAXI}/GSC observation. $^a$Exposures time are given in sec.
$^b$ Line of sight hydrogen column density is in 10$^{22}$ cm$^{-2}$ is given in Col. 6.}
\leftline{TCAF model fitted/derived parameters are mentioned in Cols. 7-12. Mass of the BH is frozen at 9.1 $M_{\odot}$. 
The horizontal lines seperate different spectral states.}
\leftline{$^c$ Accretion rates ($\dot{m_d}$ and $\dot{m_h}$) are in Eddington accretrion rate ($\dot{M}_{Edd}$). 
$^d$ Shock location is in Schwarzschild radius ($r_s$).}
\leftline{Gaussian model fitted parameters are mentioned in Col. 13-15. $^e$ Iron line energy (LE) and line width (LW) are given in keV.}
\leftline{Best fitted values of $\chi^2$ and degrees of freedom are mentioned in Col. 16 as $\chi^2/dof$. 
Errors are obtained using `{\tt err}' task in {\tt XSPEC} with 90\% confidence.}
}
\end{longrotatetable}

\appendix
Initially, we kept the mass of the BH free while fitting with the TCAF model. Each spectrum gives us a best-fitted
value of the mass. The variation of the mass is given in the Table A. From the variation, we estimated the mean value of 
the mass, which is 9.1~$M_{\odot}$. We kept the mass frozen at 9.1~$M_{\odot}$ and re-fitted all the spectra to get the
final result. The final result is quoted in Table 1. In Table A, we list the TCAF model fitted result when the mass of the 
BH was kept free. Note that, Table A does not contain the final result.

\begin{table*}[b]
\addtolength{\tabcolsep}{-4.50pt}
\small
\centering
\centering{\large \bf Table A}
\vskip -0.2cm
\centerline {}
\vskip 0.0cm
\begin{tabular}{lccccccc}
\hline
Day&$M_{BH}$ $^a$&$\dot{m_d}^b$&$\dot{m_h}^b$&$X_{S}$ $^c$& $R$ &$N$&$\chi^2 /dof$\\
(MJD) &($M_{\odot}$ ) &  ($\dot{M}_{Edd}$) & ($\dot{M}_{Edd}$) & ($r_s$) & & \\
 (1)&  (2)  & (3)  & (4)& (5) & (6) & (7) & (8)   \\
\hline

58509.45 &~ $ 9.07^{+0.53 }_{-0.42 } $  &~ $  0.36^{+0.02 }_{-0.02 } $  &~ $  0.41^{+0.03 } _{-0.02 } $&~ $  274^{+7 }_{-3 } $  &~ $  3.06^{+0.12 }_{-0.13 } $  &~$ 266 ^{+22 }_{-24 } $&  314/286 \\ 
58509.51 &~ $ 8.82^{+0.62 }_{-0.76 } $  &~ $  0.46^{+0.03 }_{-0.02 } $  &~ $  0.52^{+0.02 } _{-0.02 } $&~ $  248^{+5 }_{-9 } $  &~ $  2.49^{+0.12 }_{-0.11 } $  &~$ 411 ^{+25 }_{-22 } $&  58 /52  \\ 
58510.05 &~ $ 9.16^{+0.45 }_{-0.61 } $  &~ $  0.55^{+0.06 }_{-0.04 } $  &~ $  0.54^{+0.03 } _{-0.03 } $&~ $  224^{+4 }_{-7 } $  &~ $  1.80^{+0.12 }_{-0.11 } $  &~$ 292 ^{+16 }_{-20 } $&  317/264 \\ 
58511.58 &~ $ 9.61^{+0.91 }_{-0.66 } $  &~ $  0.57^{+0.04 }_{-0.05 } $  &~ $  0.56^{+0.04 } _{-0.05 } $&~ $  215^{+6 }_{-3 } $  &~ $  1.51^{+0.02 }_{-0.03 } $  &~$ 195 ^{+21 }_{-18 } $&  299/261 \\ 
58512.43 &~ $ 9.13^{+0.49 }_{-0.38 } $  &~ $  0.62^{+0.05 }_{-0.07 } $  &~ $  0.58^{+0.03 } _{-0.01 } $&~ $  208^{+6 }_{-8 } $  &~ $  1.48^{+0.13 }_{-0.16 } $  &~$ 336 ^{+34 }_{-42 } $&  327/265 \\ 
58513.11 &~ $ 8.75^{+0.83 }_{-0.72 } $  &~ $  0.64^{+0.04 }_{-0.05 } $  &~ $  0.60^{+0.05 } _{-0.04 } $&~ $  205^{+4 }_{-5 } $  &~ $  1.40^{+0.09 }_{-0.11 } $  &~$ 163 ^{+14 }_{-20 } $&  725/582 \\ 
58515.75 &~ $ 9.38^{+0.37 }_{-0.55 } $  &~ $  0.72^{+0.07 }_{-0.08 } $  &~ $  0.59^{+0.02 } _{-0.03 } $&~ $  187^{+4 }_{-5 } $  &~ $  1.21^{+0.09 }_{-0.04 } $  &~$ 861 ^{+57 }_{-40 } $&  358/296 \\ 
\hline
58517.67 &~ $ 9.68^{+0.72 }_{-0.87 } $  &~ $  0.73^{+0.08 }_{-0.05 } $  &~ $  0.65^{+0.03 } _{-0.05 } $&~ $  166^{+4 }_{-5 } $  &~ $  1.14^{+0.10 }_{-0.09 } $  &~$ 284 ^{+30 }_{-44 } $&  223/180 \\ 
\hline
58521.07 &~ $ 9.64^{+0.71 }_{-0.88 } $  &~ $  0.99^{+0.10 }_{-0.09 } $  &~ $  0.73^{+0.04 } _{-0.02 } $&~ $  94 ^{+3 }_{-4 } $  &~ $  1.06^{+0.05 }_{-0.04 } $  &~$ 125 ^{+20 }_{-29 } $&  137/127 \\ 
58523.00 &~ $ 9.10^{+0.73 }_{-0.57 } $  &~ $  1.34^{+0.08 }_{-0.08 } $  &~ $  0.71^{+0.05 } _{-0.03 } $&~ $  61 ^{+3 }_{-5 } $  &~ $  1.10^{+0.07 }_{-0.03 } $  &~$ 164 ^{+16 }_{-24 } $&  128/129 \\ 
\hline
58524.53 &~ $ 9.37^{+0.52 }_{-0.62 } $  &~ $  1.60^{+0.10 }_{-0.09 } $  &~ $  0.61^{+0.04 } _{-0.05 } $&~ $  51 ^{+3 }_{-4 } $  &~ $  1.05^{+0.05 }_{-0.06 } $  &~$ 307 ^{+43 }_{-53 } $&  235/208 \\ 
58527.44 &~ $ 8.44^{+0.71 }_{-0.47 } $  &~ $  1.35^{+0.10 }_{-0.15 } $  &~ $  0.49^{+0.04 } _{-0.05 } $&~ $  65 ^{+5 }_{-4 } $  &~ $  1.15^{+0.14 }_{-0.14 } $  &~$ 79  ^{+12 }_{-14 } $&  125/132 \\ 
58530.29 &~ $ 9.72^{+0.96 }_{-0.81 } $  &~ $  1.43^{+0.05 }_{-0.13 } $  &~ $  0.41^{+0.06 } _{-0.05 } $&~ $  71 ^{+2 }_{-3 } $  &~ $  1.23^{+0.05 }_{-0.10 } $  &~$ 86  ^{+13 }_{-12 } $&  142/135 \\ 
58531.35 &~ $ 9.23^{+0.65 }_{-0.75 } $  &~ $  1.27^{+0.12 }_{-0.14 } $  &~ $  0.45^{+0.02 } _{-0.05 } $&~ $  83 ^{+5 }_{-5 } $  &~ $  1.24^{+0.09 }_{-0.15 } $  &~$ 43  ^{+13 }_{-9  } $&  130/135 \\ 
\hline
58533.54 &~ $ 9.27^{+0.87 }_{-0.63 } $  &~ $  1.10^{+0.06 }_{-0.08 } $  &~ $  0.45^{+0.05 } _{-0.04 } $&~ $  95 ^{+4 }_{-3 } $  &~ $  1.33^{+0.11 }_{-0.09 } $  &~$ 58  ^{+12 }_{-14 } $&  253/223 \\ 
58535.80 &~ $ 8.50^{+0.92 }_{-0.61 } $  &~ $  1.08^{+0.10 }_{-0.12 } $  &~ $  0.43^{+0.02 } _{-0.02 } $&~ $  98 ^{+5 }_{-4 } $  &~ $  1.41^{+0.08 }_{-0.12 } $  &~$ 121 ^{+11 }_{-11 } $&  267/223 \\  
58550.49 &~ $ 8.91^{+0.71 }_{-0.60 } $  &~ $  0.78^{+0.07 }_{-0.06 } $  &~ $  0.30^{+0.03 } _{-0.02 } $&~ $  133^{+4 }_{-4 } $  &~ $  1.37^{+0.05 }_{-0.10 } $  &~$ 302 ^{+40 }_{-46 } $&  268/211 \\  
58553.87 &~ $ 9.10^{+0.62 }_{-0.49 } $  &~ $  0.64^{+0.03 }_{-0.06 } $  &~ $  0.32^{+0.02 } _{-0.01 } $&~ $  137^{+2 }_{-6 } $  &~ $  1.44^{+0.09 }_{-0.10 } $  &~$ 476 ^{+40 }_{-43 } $&  251/217 \\  
58559.13 &~ $ 8.73^{+0.88 }_{-0.50 } $  &~ $  0.50^{+0.02 }_{-0.03 } $  &~ $  0.29^{+0.01 } _{-0.04 } $&~ $  147^{+5 }_{-6 } $  &~ $  1.54^{+0.11 }_{-0.11 } $  &~$ 824 ^{+46 }_{-61 } $&  959/890 \\  
58562.04 &~ $ 8.99^{+0.79 }_{-0.64 } $  &~ $  0.41^{+0.04 }_{-0.05 } $  &~ $  0.22^{+0.02 } _{-0.02 } $&~ $  150^{+5 }_{-7 } $  &~ $  1.47^{+0.10 }_{-0.05 } $  &~$ 323 ^{+41 }_{-32 } $& 1024/879 \\ 
58574.38 &~ $ 9.38^{+0.55 }_{-0.33 } $  &~ $  0.30^{+0.03 }_{-0.02 } $  &~ $  0.16^{+0.01 } _{-0.01 } $&~ $  155^{+5 }_{-6 } $  &~ $  1.55^{+0.13 }_{-0.12 } $  &~$ 95  ^{+19 }_{-14 } $&  901/884 \\ 
58577.25 &~ $ 9.00^{+0.68 }_{-0.49 } $  &~ $  0.26^{+0.01 }_{-0.03 } $  &~ $  0.17^{+0.02 } _{-0.02 } $&~ $  165^{+7 }_{-4 } $  &~ $  1.83^{+0.13 }_{-0.12 } $  &~$ 299 ^{+25 }_{-24 } $&  282/266 \\ 
58585.20 &~ $ 9.25^{+0.22 }_{-0.33 } $  &~ $  0.15^{+0.01 }_{-0.02 } $  &~ $  0.13^{+0.02 } _{-0.02 } $&~ $  164^{+7 }_{-4 } $  &~ $  2.00^{+0.09 }_{-0.11 } $  &~$ 391 ^{+41 }_{-35 } $& 1171/891 \\ 
58588.72 &~ $ 8.48^{+0.59 }_{-0.52 } $  &~ $  0.12^{+0.01 }_{-0.01 } $  &~ $  0.12^{+0.01 } _{-0.01 } $&~ $  169^{+7 }_{-8 } $  &~ $  2.25^{+0.11 }_{-0.13 } $  &~$ 209 ^{+12 }_{-22 } $&  988/891 \\  
\hline
58597.04 &~ $ 8.62^{+0.75 }_{-0.67 } $  &~ $  0.13^{+0.01 }_{-0.02 } $  &~ $  0.11^{+0.00 } _{-0.01 } $&~ $  184^{+5 }_{-6 } $  &~ $  2.30^{+0.12 }_{-0.14 } $  &~$ 285 ^{+33 }_{-25 } $&  965/899 \\ 
\hline
58615.97 &~ $ 8.70^{+0.71 }_{-0.53 } $  &~ $  0.07^{+0.01 }_{-0.01 } $  &~ $  0.09^{+0.01 } _{-0.01 } $&~ $  194^{+5 }_{-9 } $  &~ $  2.51^{+0.14 }_{-0.12 } $  &~$ 125 ^{+13 }_{-15 } $&  689/916 \\ 
58618.74 &~ $ 8.94^{+0.54 }_{-0.47 } $  &~ $  0.07^{+0.02 }_{-0.01 } $  &~ $  0.09^{+0.01 } _{-0.01 } $&~ $  180^{+6 }_{-7 } $  &~ $  2.94^{+0.12 }_{-0.15 } $  &~$ 282 ^{+26 }_{-41 } $&  703/891 \\  
\hline
\end{tabular}
\noindent{
\leftline{TCAF model fitted/derived parameters are mentioned in Cols. 2-7.}
\leftline{TCAF model fitted mass is given in $M_{\odot}$.}
\leftline{$^b$ Accretion rates ($\dot{m_d}$ and $\dot{m_h}$) are in Eddington accretrion rate ($\dot{M}_{Edd}$). }
\leftline{$^c$ Shock location is in Schwarzschild radius ($r_s$).}
\leftline{Best fitted values of $\chi^2$ and degrees of freedom are mentioned in Col. 8 as $\chi^2/dof$.}
\leftline{Errors are obtained using `{\tt err}' task in {\tt XSPEC} with 90 \% confidence.}
\leftline{The horizontal lines seperate different spectral states.}
}
\end{table*}

\end{document}